\documentclass[9pt, twoside]{article}

\usepackage{arxiv}
\usepackage{a4wide}
\usepackage[utf8]{inputenc} 
\usepackage[T1]{fontenc}
\usepackage[bookmarks=false,colorlinks=true,urlcolor=black,linkcolor=blue,citecolor=black]{hyperref}
\usepackage{url}
\usepackage{booktabs}
\usepackage{longtable}
\usepackage{amsfonts,amsfonts,amscd,amsxtra,mathrsfs,nicefrac,pifont}
\usepackage{bold-extra}
\usepackage{microtype}
\usepackage[capitalize]{cleveref}
\usepackage{graphicx}
\usepackage[rightcaption]{sidecap}
\usepackage{xcolor}
\usepackage{natbib}
\usepackage{doi}

\newcommand{\mbf}[1]{\mathbf{#1}}

\DeclareMathOperator{\PP}{\mathbb{P}}
\DeclareMathOperator{\EE}{\mathbb{E}}
\DeclareMathOperator{\red}{red}
\DeclareMathOperator{\unq}{unq}
\DeclareMathOperator{\syn}{syn}
\DeclareMathOperator{\res}{res}

\DeclareMathOperator{\sx}{sx}
\setcitestyle{authoryear,open={(},close={)}}
\title{A General Framework for Interpretable Neural Learning based on Local Information-Theoretic Goal Functions}

\usepackage{authblk}

\newbox{\orcid}\sbox{\orcid}{\includegraphics[scale=0.06]{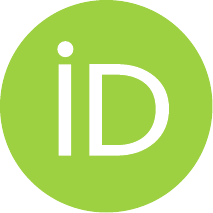}} 
\author[a,e]{%
	\href{https://orcid.org/0000-0002-3581-8262}{\usebox{\orcid}\hspace{1mm}Abdullah Makkeh\thanks{These authors contributed equally to the work}\thanks{To whom correspondence should be addressed. E-mail: \{abdullah.alimakkeh, michael.wibral\}@uni-goettingen.de; marcel.graetz@research.fchampalimaud.org; andreas.schneider@ds.mpg.de}}%
}
\author[a,b,c]{%
\href{https://orcid.org/0000-0001-7561-3408}{\usebox{\orcid}\hspace{1mm}Marcel Graetz$^{*}$$^{\dagger}$}%
}
\author[d,e]{%
\href{https://orcid.org/0009-0003-4105-1082}{\usebox{\orcid}\hspace{1mm}Andreas C.\ Schneider$^{*}$$^{\dagger}$}%
}
\author[a,e]{%
\href{https://orcid.org/0009-0009-4232-9342}{\usebox{\orcid}\hspace{1mm}David A.\ Ehrlich}
}
\author[d,e,f]{%
\href{https://orcid.org/0000-0001-8905-5873}{\usebox{\orcid}\hspace{1mm}Viola Priesemann}
}
\author[a]{%
\href{https://orcid.org/0000-0001-8010-5862}{\usebox{\orcid}\hspace{1mm}Michael Wibral$^{\dagger}$}
}
\affil[a]{Department of Data-driven Analysis of Biological Networks\\
	G\"{o}ttingen Campus Institute for Dynamics of Biological Networks\\
      University of G\"{o}ttingen\\
      G\"{o}ttingen, Germany}
\affil[b]{Department of Chemistry and Applied Biosciences\\
      ETH Zurich\\
      Zurich, Switzerland}
\affil[c]{Present address: Champalimaud Neuroscience Programme\\
Champalimaud Centre for the Unknown\\
Lisbon, Portugal}
\affil[d]{University of G\"{o}ttingen\\
      G\"{o}ttingen, Germany}
\affil[e]{Complex Systems Theory\\
      Max Planck Institute for Dynamics and Self-Organization\\
      G\"{o}ttingen, Germany}
\affil[f]{Cluster of Excellence “Multiscale Bioimaging: from Molecular Machines to Networks of Excitable Cells” (MBExC)\\
	University of Göttingen\\
 	Göttingen, Germany}

\renewcommand{\headeright}{Published in PNAS}
\renewcommand{\undertitle}{Published in PNAS~\url{https://doi.org/10.1073/pnas.2408125122}}
\renewcommand{\shorttitle}{Makkeh el al.\ A General Framework for Interpretable Neural Learning }

\hypersetup{
pdftitle={A General Framework for Interpretable Neural Learning based on Local Information-Theoretic Goal Functions},
pdfsubject={cs.IT, cs.LG, cs.NE},
pdfauthor={Abdullah Makkeh, Marcel Graetz, Andreas C.\ Schneider, David A.\ Ehrlich, Viola Priesemann, Michael Wibral},
pdfkeywords={information theory, partial information decomposition, neural networks, local learning},
}

\begin{document}
\maketitle

\begin{abstract}
	Despite the impressive performance of biological and artificial networks, an intuitive understanding of how their local learning dynamics contribute to network-level task solutions remains a challenge to this date. Efforts to bring learning to a more local scale indeed lead to valuable insights, however, a general constructive approach to describe local learning goals that is both interpretable and adaptable across diverse tasks is still missing. We have previously formulated a local information processing goal that is highly adaptable and interpretable for a model neuron with compartmental structure. Building on recent advances in Partial Information Decomposition (PID), we here derive a corresponding parametric local learning rule, which allows us to introduce 'infomorphic' neural networks. We demonstrate the versatility of these networks to perform tasks from supervised, unsupervised and memory learning. By leveraging the interpretable nature of the PID framework, infomorphic networks represent a valuable tool to advance our understanding of the intricate structure of local learning.
\end{abstract}


\section{Introduction}
Both biological neural networks (BNNs) and artificial neural networks (ANNs) are capable of solving a variety of complex tasks, thanks to their interconnected structure comprising a large number of similar computational elements. The human neocortex employs a variety of neuron types organized into canonical, repeating microcircuits that show high functional flexibility~\citep{creutzfeldt1977generality, lodato2015generating, harris2015neocortical}, similar to how ANNs utilize relatively simple processing units arranged in repetitive structures~\citep{montesinos2022fundamentals}. This structural repetition combined with functional flexibility enables both types of networks to scale drastically in size and complexity.
Given the high intrinsic complexity of these networks, achieving an interpretable understanding of how local computational elements coordinate to address global tasks is challenging and remains an ongoing focus of intense research for both BNNs~\citep{dumont2023biologically, quiroga2012concept} and ANNs~\citep{rauker2023toward,angelov2021explainable}. Despite advances towards mechanistic interpretability of the inner local computational structures that emerge through learning~\citep{goh2021multimodal, tan2022interpretable}, the insights gained from post-hoc approaches are specific to the data and network architecture, limiting their generality.

To foster a more general understanding of the local structures in neural networks, a data-independent description of the local algorithm is favorable. Such a description can be achieved through identifying a local optimization goal or learning rule, which prioritizes the learning process over the resulting representation. Traditionally, local learning has largely been formulated from two general perspectives:
On one hand, the experimental study of BNNs has revealed activity-dependent changes of synaptic strengths. This has led researchers to propose a remarkable variety of local learning rules~\citep{konorski1948conditioned, oja1982simplified, bienenstock1982theory, bi1998synaptic, isomura2016local}, most of which focus on biologically plausible mechanisms and require only locally available information. Despite these efforts, building large and powerful networks using only these mechanistic local learning rules has proven challenging~\citep{jeon2023distinctive}.
On the other hand, local learning in ANNs typically emerges implicitly by setting network-wide goal functions to satisfy global task requirements and then optimizing the network parameters via non-local gradient optimization. Such an approach hinders insights at the local scale, as the description of neuron function remains purely arithmetic.
Nonetheless, efforts have recently intensified in developing learning rules that are both local, i.e.~relying only on information that is available at the site, and show potential for scaling to larger, more capable networks~\citep{lillicrap2020backpropagation, richards2019deep,jeon2023distinctive}. This includes learning rules based on concepts from contrastive learning~\citep{illing2021local, ahamed2023forward}, predictive coding~\citep{mikulasch2023error, sacramento2018dendritic, millidge2022predictive}, local information maximization~\citep{lowe2019putting, isomura2016local} and many others~\citep{launay2020direct, hinton2022forward,hoier2023dual, lee2015difference, nokland2019training, lassig2023bio}.

Despite this large variety of fruitful efforts towards more local forms of learning, most existing approaches are limited to specific learning paradigms and implementations. What seems to be missing is a unifying framework to describe local learning goals -- general enough to be applied across learning paradigms, datasets and implementations, while still being interpretable. A promising starting point for developing such a framework from first principles is information theory~\citep{wibral2017goal-function,kay1994information, kay1997activation,kay1999neural,kay2011coherent, koren2023computational}. From an information-theoretic perspective, the local computational elements in a neural network can be interpreted as information channels that convert incoming signals into outgoing activity~\citep{wibral2015bits}, with the conversion being specified by their synaptic weights. Previous research has demonstrated the feasibility and potential of a framework of local information-theoretic goal functions based on a decomposition of the output information of the individual computational elements ~\citep{kay1994information,kay2011coherent}. 

However, since classical information theory is constructed from an information channel (simple input-to-output) perspective, it is fundamentally limited in its ability to describe all facets of information \emph{processing}:
Both the proposed biological learning in neurons and most proposed biologically plausible local learning in ANNs require at least two \textit{qualitatively different classes} of inputs to the computational element, one carrying the information to be processed and the other one carrying contextual information on how to process it (e.g.~feed-back, label, error, lateral, contrastive or reward signals -- \citealp{chereau2022circuit,schultz1998predictive, rolls1997neural,shu2003turning,manita2015top-down}).
To be able to capture the general interactions that could arise between these two classes of inputs, \cite{wibral2017goal-function} proposed a generalization and unification of existing information-theoretic local goal functions by employing the more expressive and intuitive \emph{Partial Information Decomposition} (PID). PID provides a comprehensive information-theoretic description of the complex interactions of multiple sources with respect to a target by dissecting the mutual information into unique, redundant and synergistic contributions ~\citep{williams2010nonnegative,lizier2018information,makkeh2021isx,gutknecht2021bits}\footnote{Recently, PID has also been used to describe the function of cortical neurons \citep{schulz2021gaba} and the representation of information in artificial and biological neural networks~\citep{tax2017partial,ehrlich2023a,varley2023information,ingel2022quantifying}.}.

For the case of two input classes, PID distinguishes four contributions to the overall information processing: Each class may contribute uniquely to the output, meaning they contribute information the other source does not have, they can provide redundant information or they could contribute synergistically, i.e., in a way that no input class can do alone. Taken by itself, each input class can only provide the redundant information and its unique contribution, while the synergy relies on access to both classes simultaneously (\cref{fig:neuron-model} C and D). Here, we argue that different learning paradigms require different processing of the local information from the two classes. For instance, in a supervised setting, where one class provides input data and the other provides the ground-truth labels, the intuitive goal becomes to encode in the output what is redundant between these two input classes, which enables the network to learn to extract the label information from the input signal. In general, the interpretability of these information atoms allows to intuitively identify which information processing is necessary at a local level to achieve a global task.

In this work, we derive a parametric local learning rule from a general PID-based goal function, leveraging the differentiable $i^{\sx}_\cap$ PID measure by \cite{makkeh2021isx}. We provide a proof of principle that this local learning rule enables networks consisting of compartmental neurons to solve tasks across three classic learning paradigms -- supervised, unsupervised and associative memory learning. Our work additionally shows that PID-based goals can be flexibly applied to different datasets and architectures, while being intuitively interpretable. Note that the relatively simple networks studied in this work should be considered an initial step towards larger and more capable network structures and provide evidence for the promising potential of such a general framework of interpretable local learning goals.

Below, we first explain our view of neurons as information processors with multi-class input, efficiently characterized by PID.
Based on these insights, we then introduce a compartmental neuron model and apply it to a collection of learning scenarios. We conclude with a discussion of strengths, limitations and next steps. As a side note, the neurons and networks developed in this work are termed \emph{infomorphic} -- as a portmanteau of ``information'' and ``morphous'' to indicate that they are directly shaped by the information they process.

\section{Using information theory to describe the information processing of a neuron}
\label{sec:neurons_and_infotheory}
In general, a neuron can be regarded as a Shannon information channel receiving synaptic inputs $\mathbf X$ from its afferent synapses and producing its own activity $Y$ as output. Here both $\mathbf X$ and $Y$ are modeled as random variables and their relationship can be, in the general case, stochastic. The mutual information \citep{shannon1948mathematical, cover2006elements}
\begin{equation*}
    I(Y:\mbf X) = \EE_{y, \mbf x} \log_2 p(y|\mbf x)/p(y)
\end{equation*}
then quantifies how much a neuron's output is influenced by its synaptic inputs, whereas the residual (or conditional) entropy
\begin{equation*}
    H(Y\mid \mbf X) = -\EE_{y, \mbf x} \log_2 p(y\mid\mbf x)
\end{equation*}
quantifies the amount of stochasticity in the output of the neuron that is not predictable from its inputs.
The sum of these quantities equals the total entropy or information content of the firing of the neuron
\begin{equation}
    H(Y) = I(Y: \mbf X) + H(Y\mid \mbf X).
    \label{eq:mi_plus_noise}
\end{equation}
    
\subsection{Beyond simple channels: Differentiating input classes}
\label{subsec:neurons_beyond_channels}

The picture of neurons as simple information channels has to be refined in light of the insight that different information streams into a neuron often play qualitatively different roles. In ANNs, forward-propagated signal and back-propagated gradients influence the neuron in very different ways. Similarly, biological neurons often have multiple \emph{classes} of inputs with distinct information processing characteristics~\citep{larkum2008synaptic}. An example of a biological neuron with two distinct input classes can be found in layer-5 pyramidal neurons~\citep{cajal1893nuevo}. These neurons are ubiquitous in cortex, involved in sensory, cognitive, and motor tasks, and have been hypothesized to play a role in conscious awareness~\citep{lodato2015generating,aru2020cellular}. They are typically embedded in a relatively stereotyped cortical microcircuit, at the junction of feed-forward and feed-back information streams in the cortical hierarchy~\citep{bastos2012canonical}. To process these two information streams, pyramidal neurons possess two distinct types of dendrites, the basal and apical dendrites~\citep{cajal1893nuevo}. Basal dendrites receive input from hierachically lower cortical areas and play a role in encoding the external features of the environment that are processed along the cortical hierarchy~\citep{chereau2022circuit}. Apical dendrites, in contrast, receive contextual input from higher cortical areas and have been shown to play an important role in modulating perception~\citep{takahashi2016active,takahashi2020active}. This connectivity is similar across a range of different brain areas and cognitive domains, motivating the assumption that the general function of pyramidal neurons is independent of the semantics of their input (\cref{fig:neuron-model} A and ref.~\citealp{rockel1980basic}).

The two-compartment structure of layer-5 pyramidal neurons~\citep{kording2001supervised} is consistent with many biologically plausible local learning rules in ANNs that require at least two \textit{qualitatively different classes} of inputs to the neuron, respectively carrying feedforward information to be processed and contextual information to guide this processing (e.g.~feed-back, label, error, lateral, contrastive or reward signals -- \citealp{chereau2022circuit,schultz1998predictive, rolls1997neural,shu2003turning,manita2015top-down}). Once these two input classes are explicitly established, they motivate a local learning goal.

To prepare for the mathematical representation of a neuron's unique, redundant and synergistic information, we will first reinterpret the source variable $\mbf X$ from above as being a composite variable $\mbf X = (\mbf X_R, \mbf X_C)$ of the \emph{receptive} input $\mbf X_R$, which is inspired by the input to the basal dendrites, and the \emph{contextual} input $\mbf X_C$, which is inspired by the input to the apical dendrites. Analogous to \eqref{eq:mi_plus_noise}, the total entropy of the neuron $Y$ can now be written as
\begin{equation}
    H(Y) = I(Y:\mbf X_R, \mbf X_C) + H(Y\mid \mbf X_R, \mbf X_C).
    \label{eq:h_it}
\end{equation}

The dissection of $\mbf X$ additionally allows to consider the individual channels of the receptive or contextual inputs to the target, which are characterized by the mutual information terms $I(Y:\mbf X_R)$ or $I(Y: \mbf X_C)$, respectively. Note, however, that these two channels do not simply add up to the total mutual information $I(Y:\mbf X_R, \mbf X_C)$, because the sum $I(Y:\mbf X_R) + I(Y:\mbf X_C)$ contains information which is redundantly present in both input classes and will be double-counted, while synergistic information which only becomes apparent if one considers $\mbf X_C$ and $\mbf X_R$ simultaneously will be overlooked (\cref{fig:neuron-model} C, \citealp{mcgill1954multivariate,cover2006elements}). By introducing the co-information
\begin{equation}
    \begin{aligned}
        I(Y:\mbf X_R:\mbf X_C) &= I(Y: \mbf X_R, \mbf X_C)\\
        &- I(Y:\mbf X_R \mid \mbf X_C) - I(Y:\mbf X_C \mid \mbf X_R),
    \end{aligned}
    \label{eq:coinfo}
\end{equation}
the decomposition in \eqref{eq:h_it} can be refined to 
\begin{equation}
    \begin{aligned}
        H(Y) &= I(Y:\mbf X_R:\mbf X_C)\\
        &+ I(Y:\mbf X_R \mid \mbf X_C) + I(Y:\mbf X_C \mid \mbf X_R)\\
        &+ H(Y\mid \mbf X_R, \mbf X_C),
    \end{aligned}
    \label{eq:h_kay}
\end{equation}
where the conditional mutual information $I(Y:\mbf X_R \mid \mbf X_C)$ captures the remaining dependence of $Y$ on $\mbf X_R$ when $\mbf X_C$ is known, and $I(Y:\mbf X_C \mid \mbf X_R)$ is defined analogously~\citep{cover2006elements}.

\cite{kay1994information} used this decomposition as the starting point to construct models of learning neurons with information theoretic objective functions. In our work, we build on this concept by exploiting the superior expressiveness provided by the framework of Partial Information Decomposition to build infomorphic neurons.

\subsection{Uncovering the information processing between different input classes using Partial Information Decomposition\label{subsec:pid}}

The above perspective of viewing a neuron as a collection of information channels still paints an incomplete picture of the information processing within a neuron because it cannot account for all the different ways in which the different information sources combine and determine the output information: While some of the information in the neuron's output activity $Y$ might be provided \emph{uniquely} by either the receptive input $\mbf X_R$ or the contextual input $\mbf X_C$, other parts might be \textit{redundantly} supplied by both of them while yet others only become available \emph{synergistically} when both sources are considered jointly~\citep{wibral2017goal-function}. Classical information theory is insufficient for this distinction as it has no concept of ``sameness'' of information: While one can compute the total amount of information in the output that is coming from each source or from both sources together using mutual information, there is no way of quantifying how much of the information contributed to the output is the same, i.e., redundantly provided by the input variables about the output~\citep{williams2010nonnegative}.

\begin{figure}[t]
    \centering
    \includegraphics[width=\textwidth]{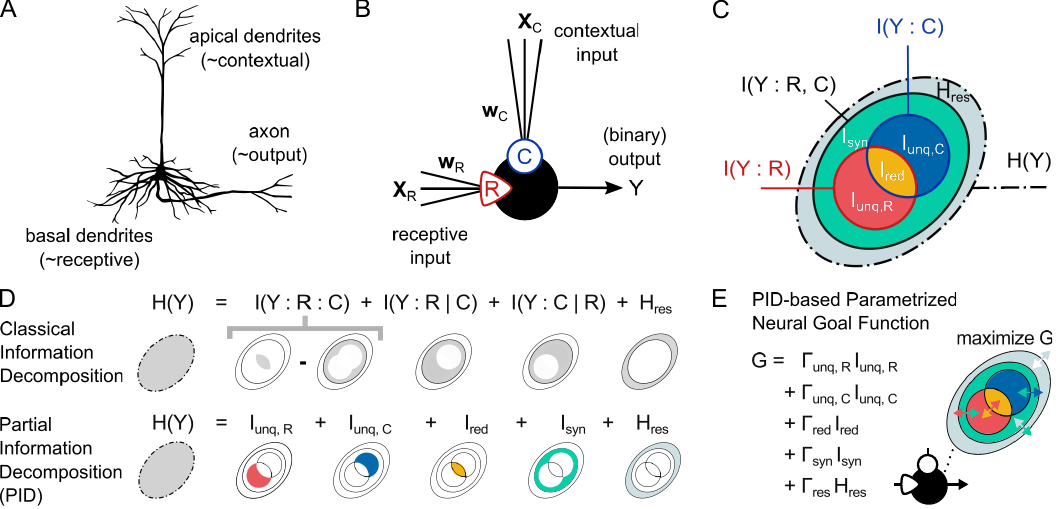}
    \caption{\textbf{The infomorphic neuron model, analogous to cortical pyramidal neurons, separately integrates two distinct classes of inputs. The neuron adjusts its synaptic weights to maximize the local goal function $G$, based on an information-theoretic decomposition of its own output information.}
    (A) Cortical pyramidal neurons with separate synaptic integration sites for basal and apical dendrites, the former driving output and the latter providing contextual modulation.
    (B) The infomorphic neuron, modeled after these neurons, is characterized by two functionally distinct sets of inputs that are scaled by synaptic weights and added to obtain the integrated input signals $R$ (receptive) and $C$ (contextual). R and C contribute individually to the probabilities of the neuron's binary output, which are computed using an activation function $A(R,C)$ and a sigmoid transformation function.
    (C) The total Shannon output information $H(Y)$ of the neuron consists of the mutual information with the joint inputs $I(Y : R, C)$ and additional residual information $H_{\res} = H(Y \mid R, C)$ that originates directly from the stochasticity of the neuron. Using Partial Information Decomposition (PID), the joint mutual information $I(Y : R, C)$ can be further subdivided into four information contributions: (i) $I_{\red}$, the redundant information that is provided by either $R$ or $C$ individually, (ii) $I_{\unq, R}$, the unique information of $R$ that is only provided by $R$ but not by $C$, (iii) $I_{\unq, C}$, the unique information of $C$ that is only provided by $C$ but not by $R$, and (iv) $I_{\syn}$, the synergistic information that is provided by $R$ and $C$ only when taken jointly but neither by $R$ nor $C$ taken individually.
    (D) Any classical mutual-information-based decomposition can only provide a linear combination of the underlying PID quantities. (D Upper) Classical decomposition into four information contributions that formed the basis for prior work \citep{kay1994information,kay1997activation,kay1999neural, kay2011coherent}: the co-information $I(Y: R : C)$, the two conditional mutual information values $I(Y: R\mid C)$ and $I(Y: C\mid R)$, and the stochasticity-caused residual entropy $H_{\res}$. (D Lower) The five contributions that are quantified using PID. 
    (E)  The neuron's synaptic weights $\mbf w$ are optimized to maximize a goal function $G$ that is based on the Partial Information Decomposition of the neuron's overall output information $H(Y)$ and parameterised by $\mbf \Gamma = (\Gamma_{\unq, R}, \Gamma_{\unq, C}, \Gamma_{\red}, \Gamma_{\syn}, \Gamma_{\res})$.
    Panel (A) is adapted from Fabian Mikulasch's original depiction of a pyramidal neuron~\citep{mikulasch2023error} (CC0).
    }
    \label{fig:neuron-model}
\end{figure}

Dissecting the mutual information between multiple source variables and a single target variable into non-overlapping additive information \emph{atoms} is the subject of Partial Information Decomposition~\citep{williams2010nonnegative, gutknecht2021bits}. Using PID, we can subdivide the entropy $H(Y)$ into five parts (\cref{fig:neuron-model} C)
\begin{equation}
    \begin{aligned}
        H(Y) &= I_{\unq}(Y: \mbf X_R) + I_{\unq}(Y: \mbf X_C)\\
        &+ I_{\red}(Y: \mbf X_R, \mbf X_C) + I_{\syn}(Y: \mbf X_R, \mbf X_C)\\
        &+ H(Y\mid \mbf X_R, \mbf X_C),
    \end{aligned}
    \label{eq:h_pid}
\end{equation}
where $I_{\unq}(Y: \mbf X_R)$ and $I_{\unq}(Y: \mbf X_C)$ are the unique information atoms of the receptive and contextual inputs, respectively, $I_{\red}(Y: \mbf X_R, \mbf X_C)$ refers to the redundant (shared) information,  and $I_{\syn}(Y: \mbf X_R, \mbf X_C)$ refers to the synergistic (complementary) information.
These four atoms can describe the information processing in $Y$ of $\mbf X_R$ and $\mbf X_C$ in versatile ways, while also having meaningful interpretations: For example, if a neuron encodes the coherent parts of its inputs, this would be reflected in a high redundant information. Alternatively, a neuron might encode the information in its receptive input $\mbf X_R$ that is specifically \emph{not} present in the contextual input $\mbf X_C$, which would translate to a high unique information contribution from $\mbf X_R$. Finally, if the neuron's output contains information which cannot be obtained from any single source alone, for instance if the output $Y$ reflected the logical ``exclusive or'' of its inputs, the synergy between the sources would be high. Overall, PID provides a decomposition framework with well-defined and intuitive interpretations for understanding a neuron's information processing.

Note that while the co-information $I(Y: \mbf X_R : \mbf X_C)$ (in \eqref{eq:coinfo}) is equal to the difference between redundant and synergistic information
\begin{equation}
    I(Y: \mbf X_R : \mbf X_C) = I_{\red}(Y: \mbf X_R, \mbf X_C) - I_{\syn}(Y: \mbf X_R, \mbf X_C),\label{eq:pided-coinfo}
\end{equation}
classical information theory provides no tool to disentangle the two components.

To analyze the information processing of a neuron, the aforementioned PID atoms need to be quantified. Note that despite their strong relation to classical information-theoretic quantities through \eqref{eq:h_pid} and \eqref{eq:pided-coinfo}, the size of the PID atoms cannot be determined from classical information-theoretic quantities alone as there are four atoms with only three equations providing constraints~\citep{williams2010nonnegative}. An additional quantity has to be defined for PID, which is typically, but not necessarily, the redundant information~\citep[and references therein]{williams2010nonnegative,lizier2018information, gutknecht2021bits}.
By now, a multitude of different measures for redundant information have been proposed, each fulfilling a number of partly mutually exclusive desiderata and drawing on concepts from different fields such as decision or game theory, e.g.~\citep{harder2013bivariate,bertschinger2014quantifying,ince2017measuring,finn2018pointwise,lizier2018information}. In this work, we use the PID measure $I^{\sx}_\cap$ defined by~\cite{makkeh2021isx} due to its analytical differentiability with respect to the underlying joint probability distribution $\PP(Y, \mbf X_R, \mbf X_C)$, allowing for optimization of the PID quantities through gradient ascent.

\section{Infomorphic Neurons}
\label{sec:infomorphic_neuron}

In a line of similar work, \cite{kay1994information} utilized the decomposition in \eqref{eq:h_kay} not as a post-hoc analysis tool, but as a parameterizable optimization goal function, extending this idea in subsequent research \citep{kay1997activation, kay1999neural, kay2011coherent}. Even before the development of their differentiable PID measure, \cite{wibral2017goal-function} envisioned a similar, but more refined neural goal function derived from the decomposition in \eqref{eq:h_pid}. In the following paragraphs we realize this idea in a neuron model closely aligned to prior work \citep{kay1994information}, which we refer to as the \emph{infomorphic neuron}, and derive analytic gradients for the PID-based goal function.

\paragraph{Multi-compartment computation.} 

Infomorphic neurons operate in discrete time and output values $Y\in\{-1,+1\}$ (referred to as ``LOW'' and ``HIGH''), in analogy to time-binned spike trains of biological neurons. Akin to the basal and apical dendrites of layer-5 pyramidal neurons, an infomorphic neuron distinguishes between two classes of input synapses, namely ``receptive'' inputs $\mbf X_R$ and ``contextual'' inputs $\mbf X_C$ (\cref{fig:neuron-model} A and B). Inspired by how the inputs of different input classes are individually aggregated in separate compartments  in these biological neurons~\citep{cajal1893nuevo}, the inputs of the two classes of the infomorphic neuron are separately combined in a weighted sum to produce the aggregate inputs $R = \mbf w_R^T \mbf X_R - w_{0,R}$ and $C = \mbf w_C^T \mbf X_C - w_{0,C}$~\citep{kay1994information}. Here, $\mbf w_R$ and $\mbf w_C$ reflect the weights associated with the receptive and contextual inputs, respectively, while $w_{0,R}$ and $w_{0,C}$ denote constant bias values. At any time step, the probability $\theta$ of a neuron to be in the HIGH state depends only on the instantiation of its aggregate inputs $r$ and $c$, as follows:
\begin{equation*}
    \theta(r, c):= \PP(Y=1\mid R=r, C=c):=\sigma(A(r,c)),
\end{equation*}
where $\sigma(\xi) = 1 / (1 + e^{-\xi})$ is a sigmoid nonlinearity, and $A$ is an additional activation function. While the activation function can in principle be chosen arbitrarily, a biology-inspired choice of $A$ may draw inspiration from layer-5 pyramidal neurons: By making the activation function be primarily dependent on the receptive inputs, one can imitate the privileged role that basal dendrites play in driving pyramidal neurons~\citep{kay1997activation}. In practice, we adapted the degree to which the contextual input influences the output, dependent on the requirements of each task. The choice of activation function will be individually motivated and discussed in the corresponding experimental sections.

\paragraph{Local learning.}

Each infomorphic neuron optimizes its local information processing by changing the two sets of weights $\mbf w_R$ and $\mbf w_C$ of its incoming (afferent) synapses. This information processing can take on very different shapes: For some tasks, optimal information processing could mean coding for coherence between the receptive and contextual inputs, while for other tasks, optimal processing might entail extracting any piece of information (e.g.\ a feature) exclusively provided by the receptive inputs that is not present in the contextual input.

\cite{kay1994information} first derived a local goal function from an information-theoretic partition of the local mutual information of a neuron. Here, we argue for a similar local goal function involving a linear combination of the five components of the output entropy of a neuron as derived from PID and first established by \cite{wibral2017goal-function}:

\begin{equation}
    \begin{split}
        G(Y: \tilde R, \tilde C) 
        &= \Gamma_{\unq,R} \, I_{\unq}(Y: \tilde R)
        + \Gamma_{\unq,C} \, I_{\unq}(Y: \tilde C)\\ 
        &+ \Gamma_{\red} \, I_{\red}(Y:\tilde R, \tilde C)
        + \Gamma_{\syn} \, I_{\syn}(Y: \tilde R, \tilde C)\\
        &+\Gamma_{\res} \, H(Y\mid \tilde R, \tilde C)\\
        &=: (\Gamma_{\unq,R}, \Gamma_{\unq,C}, \Gamma_{\red}, \Gamma_{\syn}, \Gamma_{\res})\\
        &\quad\cdot(I_{\unq,R}, I_{\unq,C}, I_{\red}, I_{\syn}, H_{\res})^T.
    \end{split}
    \label{eq:goal_fct}
\end{equation}

Here, the variables $\tilde R$ and $\tilde C$ are binned versions of the continuous-valued $R$ and $C$ inputs, necessary due to the lack of a differentiable PID measure for mixed discrete-continuous variables~\citep{ehrlich2024partial} and other conceptual difficulties of information theory in continuous networks~\citep{saxe_information_2019, goldfeld_estimating_2019}. Note that the binning procedure itself, while used in analogy to previous work~\citep{kay2011coherent,kay1994information}, is a non-differentiable operation whose gradients we do not take into account here. Future work might circumvent this problem by using parametric (e.g.~bivariate Gaussian) approximations of $p(R,C)$~\citep{kay2011coherent,kay1994information} combined with PID-estimators for mixed discrete-continuous variables. The neuron's local goal function $G$ is a linear combination of the PID atoms that is defined a-priori by choosing the parameters $\mbf \Gamma$ (\cref{fig:neuron-model} E). In the second equality we introduced a short-hand vector notation of $G$.

\paragraph{Optimizing the goal function.}

The differentiability of the $I_\cap^\mathrm{sx}$ measure allows each neuron in an infomorphic network to optimize its own goal function $G$ through gradient ascent.

The empirical gradients of $G$ with respect to the weight vectors $\mbf w_R$ and $\mbf w_C$ can be analytically derived as
\begin{equation}
    \frac{\partial \hat{G}}{\partial \mbf w_R} = \frac{1}{N} \sum_{\mbf x_R, \mbf x_C} f_{p(R,C)}^{\mbf\Gamma}(\tilde r, \tilde c) \; \left.\frac{\partial A}{\partial r}\right\rvert_{\tilde r, \tilde c} \; \mbf x_R
    \label{eq:G_wR}
\end{equation}
and
\begin{equation}
    \frac{\partial \hat{G}}{\partial \mbf w_C} = \frac{1}{N} \sum_{\mbf x_R, \mbf x_C} f_{p(R,C)}^{\mbf\Gamma}(\tilde r, \tilde c) \; \left.\frac{\partial A}{\partial c}\right\rvert_{\tilde r, \tilde c} \; \mbf x_C,
    \label{eq:G_wC}
\end{equation}
where $\hat{G}$ indicates the estimator of $G$ based on $N$ input samples in the data, and $f$ is implicitly dependent on the full probability distribution $p_{\tilde R, \tilde C}$ of $\tilde r$ and $\tilde c$ over the dataset and the explicit current values of those variables, as well as the goal parameter vector $\mbf \Gamma$. The full derivation of the gradients can be found in \cref{apx:gradients}.

In practice, we only update the network parameters after a fixed number of discrete network time steps, referred to as a mini-batch. For each mini-batch we estimate the full binned probability distribution $p_{\tilde R, \tilde C}$ from the histogram of inputs and finally conduct a single weight update. We report the number of mini-batches as the training time $t$. Instead of using mini-batches, it would also be possible due to the pointwise nature of the $i^{\sx}_\cap$ PID measure to keep running estimates with exponential forgetting of past samples, which would allow for weight updates after each network time step.

\section{Infomorphic networks encompass various learning paradigms}
\label{sec:experiments}

The parameterized information-theoretic goal function enables groups of infomorphic neurons, i.e. ``infomorphic networks'', to serve as a very general and versatile approach to learning. In the following we demonstrate their broad applicability by providing three example applications of infomorphic networks, on supervised learning, unsupervised learning and online learning of associative memories.

In correspondence to classical ANNs, infomorphic networks require choices on network topology, activation functions and goal functions, where the latter are chosen by setting the goal function hyper-parameters for each neuron. The ability to arbitrate between different local goals by setting these hyper-parameters is a major strength of our framework, and we will motivate and discuss our specific hyper-parameter choices in all three presented applications.

\subsection{Supervised learning by encoding coherence between input and label information\label{subsec:supervised}}

\begin{SCfigure*}[\sidecaptionrelwidth][t]
    \centering
    \includegraphics[]{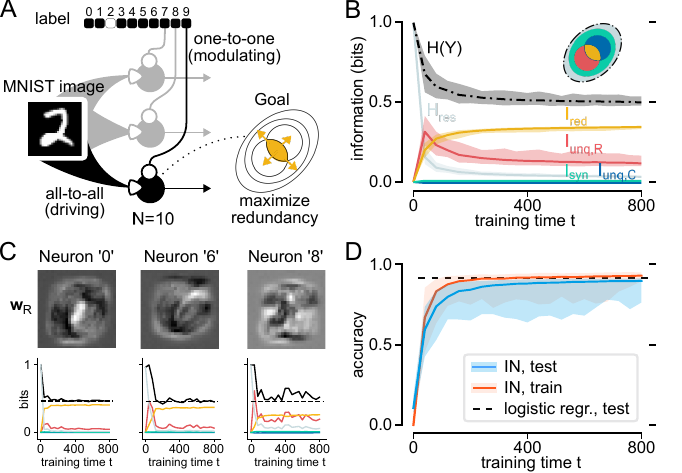}
    \caption{\textbf{Supervised learning in single-layer infomorphic networks.} By maximizing redundant information between image and label, the neurons learn to identify MNIST digits with a test accuracy comparable to logistic regression. (A) Network architecture with one-hot encoded label and 10 neurons, each receiving all 28x28 image pixels as $\mbf X_R$ and one element of the label vector as $X_C$. Activation function $A(r, c) := r(0.5 + \sigma(2rc))$ is chosen such that $c$ has only modulating effect on the binary output probabilities, in line with the label's role as context for learning. The goal of each neuron is to transmit maximal information $I_{\red}$ that is redundant between $R$ (image) and $C$ (label element), thereby learning to act as a detector of its respective digit. The learning shows best stability if the goal function sets weak incentives for additionally maximizing the unique and synergistic information: $\mbf \Gamma_{\mathrm{supervised}} = (0.1, 0.1, 1, 0.1, 0)$. (B) Information quantities averaged over all neurons, shown for 100 independent training runs. (C) Receptive fields ($\mbf w_R$) and information quantities for three sample neurons for a single training run, the dotted line indicating the expected $H(Y)$ in case of perfect classification (one-vs-all entropy of label in test data set). (D) The average training and test accuracies across 100 training runs, with test accuracy approaching that of logistic regression (reaching on average $89.7\%$ vs. $91.9\%$  for log. regr.). Note that in (B) and (D) the 95-percentile is being displayed.}
    \label{fig:supervised}
\end{SCfigure*}

We construct a single-layer infomorphic network for supervised classification of MNIST digits~\citep{lecun1998mnist}.

\paragraph{Topology and Inputs} Each of $k=10$ neurons receives the full MNIST image via a set of $28\cdot28=784$ receptive input synapses, with $\mbf X_R \in \{0,\frac{1}{255},\dots,1\}^{28\times 28}$, and a single element of a one-hot label vector as contextual input, with $X_C \in \{-1,1\}$ (\cref{fig:supervised} A). In this setup, each neuron becomes a one-vs-all classifier of its assigned digit.

\paragraph{Goal Functions} Viewed through the lens of PID, supervised learning requires extracting from input data the \emph{same} information that is contained in the label. This is achieved if each neuron's output information is redundantly determined by its two input classes, motivating the goal function $G = I_{\red}$. In practice, weak incentives for the other PID quantities $\mbf \Gamma_{\mathrm{supervised}} = (\Gamma_{\unq, R}, \Gamma_{\unq, C}, \Gamma_{\red}, \Gamma_{\syn}, \Gamma_{\res}) = (0.1, 0.1, 1, 0.1, 0)$ improve performance and stability of learning by preventing neurons from going silent, i.e., always outputting the same value.

\paragraph{Activation Functions} To ensure that the receptive inputs are strong enough to drive the neurons during the test phase when the label is missing (all label input $x_C = 0$), we set the activation function to $A_\sigma(r,c):=r(0.5 + \sigma(2rc))$. This makes the binary output probabilities mostly dependent on the receptive input and only weakly modulated by the contextual information, rendering the label input a teacher signal that strongly influences learning but hardly the dynamics.

\paragraph{Protocol} In the training phase, we present the MNIST images and labels sequentially in random order, with a weight update after each mini-batch (\cref{apx:params-supervised} for all chosen training parameters). In the test phase, we present previously unseen MNIST images and set the contextual input to $x_C = 0$, calculating winner-take-all classification accuracy. Note that instead of calculating $\PP(Y\mid r, c=0)$, an optimal predictor would marginalize over $\PP(c_\mathrm{train})$ to estimate $\PP(Y\mid r)\approx \sum_{c_\mathrm{train}}\PP(Y\mid r, c_\mathrm{train})\PP(c_\mathrm{train})$. However, this would require running each test input twice for the two different labels, and performing computations that infomorphic neurons do not implement. Fortunately, due to the merely modulating role of the contextual input, the test accuracy of both approaches is virtually identical (Appendix \cref{fig:mnist-optimal-accuracy}).

\paragraph{Performance and Outcome} The infomorphic networks reach an average test accuracy of $89.7\,\%$ (\cref{fig:supervised} D), slightly lower than the $91.9\,\%$ we find for multinomial logistic regression. Indeed, logistic regression upper-bounds the network performance, because by setting $x_C = 0$ in the test phase, the activation function simplifies to $A(r,c=0)=r$ and the firing probability becomes $\theta = \sigma(\mbf w_R\cdot \mbf x_R - w_{0,R})$, identical to logistic regression.
The receptive weights $\mbf w_R$ of individual neurons after training, plotted as receptive fields, visually reveal their assigned digit and qualitatively match the corresponding receptive fields found in logistic regression (\cref{fig:supervised} C and Appendix \cref{fig:mnist-receptive-fields}).

\paragraph{Information Dynamics} Analyzing the information atoms of individual neurons over the course of training, we find an expected increase in redundant information $I_{\red}$ (\cref{fig:supervised} B and C). This increase is less pronounced in neurons corresponding to digits that are more likely to be confused (Appendix \cref{fig:mnist-pid}). For these neurons we also find higher unique information from the receptive input $I_{\unq, R}$, indicating that they are still encoding image information that is not present in their label. Additionally, the average output entropy $H(Y)$ of the neurons decreases and approaches the average entropy of the one-hot label encoding (label $k$ present vs. absent) of $0.47$ bits, while the residual entropy $H(Y | R, C)$ decreases fast, reflecting a decrease in the neurons' stochasticity.

\begin{SCfigure*}[\sidecaptionrelwidth][t]
	\centering
    	\includegraphics[]{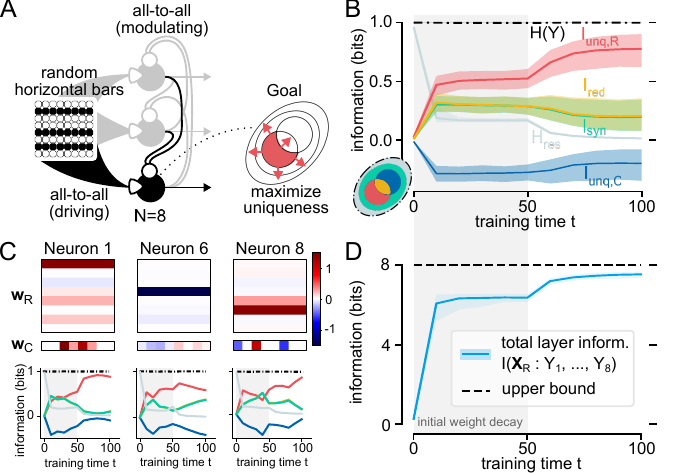}
	\caption{\textbf{Unsupervised feature learning in recurrent infomorphic networks.} By maximizing unique information with respect to all other neurons, the neurons self-organize to create a highly informative representation of the input. (A) Network architecture for unsupervised learning with 8 neurons, each receiving 8x8 binary pixel inputs as $\mbf X_R$ and output of all other neurons as $\mbf X_C$. The image consists of 8 independent horizontal bars, randomly appearing with $p=0.5$. Activation function $A(r, c) := r(0.5 + \sigma(2rc))$ is chosen such that $c$ has only modulating effect on the binary output probabilities, which leads to recurrent connections mainly acting as context for learning. The goal of each neuron is set to maximize unique information $I_{\unq, R}$ of its own receptive input $R$ with respect to the output of all other neurons, received as contextual input $C$: ${\mbf \Gamma_{\mathrm{unsupervised}} = (1, 0, 0, 0, 0)}$. (B) Information quantities averaged over all neurons for 300 independent training runs, showing two-phase training for feature competition and stabilization. (C) Receptive and contextual fields ($\mbf w_R$, $\mbf w_C$) and information quantities of three sample neurons for a single training run. (D) Mutual information $I(\mbf X_R: Y_1, ..., Y_8)$ between all neurons' outputs and input image, approaching full encoding capacity and input information content of 8 bits. Note that in (B) and (D) the 95-percentile is being displayed.}
	\label{fig:unsupervised}
\end{SCfigure*}

\subsection{\label{subsec:unsupervised}Unsupervised learning of independent features by maximizing each neuron's unique information about the stimulus}
We construct a very simple data compression task that requires recurrent communication between neurons.

\paragraph{Topology and Inputs} Each of the $k=8$ neurons receives $8\times8$-pixel binary images as receptive inputs, $\mbf X_R \in \{-1,1\}^{8\times 8}$, with each image containing 8 horizontal bars appearing independently with probability $p=0.5$ (\cref{fig:unsupervised} A). As contextual input, each neuron receives the activity of all other neurons in the previous time step, with $\mbf X_C \in \{-1,1\}^{7}$.

\paragraph{Goal Functions} The network-level goal is to encode all 8 bits of the information provided by the image distribution, distributed over the neurons. This can be achieved if each neuron encodes one full bit of image information that is not already encoded by the other neurons, and motivates a goal function maximizing for the conditional mutual information $G = I(Y:R|C) = I_{\unq, R} + I_{\syn}$. In order to encourage the network to explicitly disentangle the contributions of each neuron, we chose to only encourage the unique information of the receptive input: ${\mbf \Gamma_{\mathrm{unsupervised}} = (\Gamma_{\unq, R}, \Gamma_{\unq, C}, \Gamma_{\red}, \Gamma_{\syn}, \Gamma_{\res}) = (1, 0, 0, 0, 0)}$.

\paragraph{Activation Functions} To avoid temporal oscillations in the network, we chose the same activation function $A_\sigma(r,c):=r(0.5 + \sigma(2rc))$ as in the supervised context, making the recurrent connections relevant for learning, but less so for the dynamics.

\paragraph{Protocol} We sequentially present randomly sampled images containing between 0 and 8 bars. Due to the time delay in recurrent connections, presentation of a new image introduces a mismatch between the receptive and contextual inputs of all neurons. We compensate for the resulting noise by presenting each image for 8 successive time steps.

Early in training, neurons compete for which information to encode. Two neurons choosing to encode for the same bar leads to a local optimum where both neurons try to increase their receptive weights to reduce stochasticity, however cannot obtain high unique information. To avoid this local maximum and prolong the critical initial phase of high stochasticity and competition, we introduce a strong weight decay (linear down-scaling of all receptive weights after each time step) during the first half of training (\cref{apx:params-unsupervised} for all chosen training parameters).

\paragraph{Performance and Outcome} Over the course of training almost all neurons learn to encode mutually different individual bars (\cref{fig:unsupervised} C). Rare encoding errors occur exclusively when two neurons encode the same bar (Appendix \cref{fig:bars-encoding-all-nets}). Correspondingly, the total mutual information of the layer $I(\mbf X_R: Y_1,\ldots, Y_8)$ approaches the entropy of the data set, indicating successful compression of the receptive input information (\cref{fig:unsupervised} D).

\paragraph{Information Dynamics} The average unique receptive information of the neurons $I_{\unq,R}$ converges to 0.77 bits, with the average conditional mutual information $I(Y:R|C) = I_{\unq, R} + I_{\syn}$ reaching close to 1.0 bits (\cref{fig:unsupervised} B and C). This suboptimal result is mostly attributable to individual neurons not having fully converged onto their chosen bar, and to the weak contextual cross-talk between neurons (Appendix \cref{fig:bars-succ-pid-weights}). However, the above-mentioned rare encoding errors, i.e., two neurons encoding the same bar, show a strikingly different signature of low unique information $I_{\unq,R}$ and high redundancy $I_{\red}$ (Appendix \cref{fig:bars-unsucc-pid-weights}).

\subsection{Online associative memory learning by maximizing the local coherence between network firing and external input\label{subsec:memory}}
We construct an (auto-)associative memory network, similar to the Hopfield network~\citep{hopfield1982neural}, with a novel infomorphic online learning rule.

\begin{SCfigure*}[\sidecaptionrelwidth][t]
	\centering
	\includegraphics[]{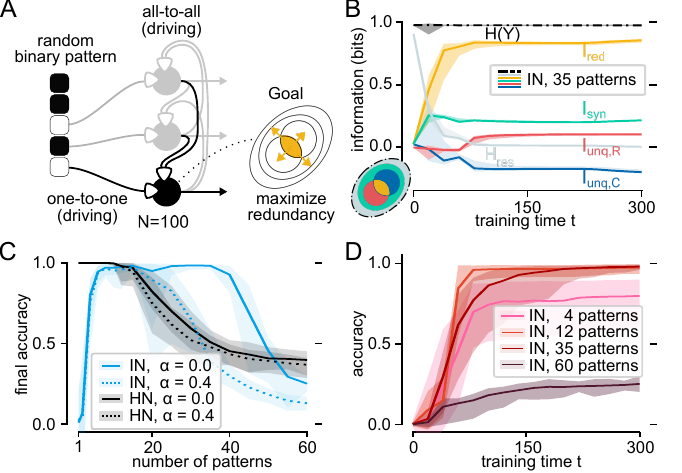}
	\caption{\textbf{Associative memory learning in recurrent infomorphic networks.} By maximizing redundant information between external input and output of all other neurons, the neurons learn to memorize a maximum number of patterns exceeding that of classical Hopfield networks. (A) Network architecture for memorizing binary patterns with 100 recurrently connected neurons, each receiving single element of target patterns as $X_R$ and output of all other neurons as $\mbf X_C$. Activation function $A(r,c)$ is chosen such that both $r$ and $c$ can directly drive the neuron's output probabilities, to enable an active recovery of patterns based only on recurrent connections at test time (\cref{subsec:memory}). The goal of each neuron is to learn to predict their respective element of the pattern based on activity of other neurons, which can be done by maximizing the redundant information in the output. Similar to the supervised example, the goal function used here includes additional, weak incentives for maximizing unique and synergistic PID contributions: ${\mbf \Gamma_{\mathrm{memory}} = (0.1, 0.1, 1, 0.1, 0)}$. (B) Information quantities averaged over all neurons, for 25 independent runs trained on 35 memory patterns. (C) Final accuracy of infomorphic networks and classical Hopfield networks over different numbers of patterns on the horizontal axis. Shown are the results for two noise levels used for testing the recovery of memorized patterns. (D) Accuracy over training for all 25 runs for different numbers of patterns. Note that in (B), (C) and (D) the 95-percentile is being displayed.}
    \label{fig:memory}
\end{SCfigure*}

\paragraph{Topology and Inputs} Each of $k=100$ neurons receives a single element of a ($p=0.5$)-sparse memory vector as receptive input, with $X_R \in \{-1,1\}$, and the activity of all other neurons in the previous time step as contextual input, with $\mbf X_C \in \{-1,1\}^{99}$ (\cref{fig:memory} A).

\paragraph{Goal Functions} The network-level goal is to align with any external input, when present, and over time inscribe it as an attractor, i.e., a memory, into the recurrent dynamics. The external input thus functions as both the memory cue and the teaching signal, depending on the duration of presentation. Learning a memory pattern upon repeated presentation implies that the recurrent contextual inputs learn to align with the external receptive inputs by providing the \emph{same} information about the firing of a neuron as the external inputs, motivating the goal function $G = I_{\red}$. As in the supervised experiment, weak incentives for the other PID quantities $\mbf \Gamma_{\mathrm{memory}} = (\Gamma_{\unq, R}, \Gamma_{\unq, C}, \Gamma_{\red}, \Gamma_{\syn}, \Gamma_{\res}) = (0.1, 0.1, 1, 0.1, 0)$ improve performance by preventing neurons from going silent.

\paragraph{Activation Functions} In the absence of a receptive input the neurons should be driven by the contextual synapses, while receptive input, if present, should overrule this recurrent drive and force the neurons into a new firing pattern.
As a consequence, each input class individually needs to be able to drive the neuron. To make both inputs driving and encourage high weights, we choose the symmetric activation function
\begin{equation*}
A(r,c) = \frac{r^8 \mathrm{sign}(r) + c^8 \mathrm{sign}(c)}{r^8+c^8} \cdot (r^8+c^8)^{1/8}
\end{equation*}
that is monotonic in both $r$ and $c$ and aligns with the positive and negative 8-norm for $r,c>0$ and $r,c<0$, respectively (\cref{apx:params-memory} for all chosen training parameters).

\paragraph{Protocol} In the training phase, we sequentially presented a set of memory patterns to the network in random order. As in the unsupervised experiments, to compensate for the time delay of the recurrent connections, we presented each memory pattern for 8 consecutive time steps. Note that structured sequences of patterns presented for only one time step lead to hetero-associative memory formation (results not reported here, but in~\citep{graetzthesis2021}).

In the test phase, we present noise-corrupted memory patterns for a single time step, with the noise level $0\leq\alpha\leq 1$ indicating the fraction of pattern elements set at random. After presentation, we set the receptive input of all neurons to $x_R = 0$ and assess retrieval accuracy by computing the cosine similarity between the network state and the noiseless memory pattern after 20 time steps (\cref{fig:memory}D). We define the noise-dependent capacity of the network as the highest number of trained memory patterns such that the average retrieval accuracy exceeds $95\%$.

\paragraph{Performance and Outcome} For noiseless initialization, infomorphic networks attain a capacity of 35 patterns per 100 neurons, which significantly exceeds the limit of 14 patterns in classical Hopfield networks~\citep{hopfield1982neural}. Note that unlike Hopfield networks our readout includes no binarization but remains stochastic, thus likely even limiting the measured capacity in the direct comparison. Furthermore, infomorphic networks outperform Hopfield networks up to a noise level of $\alpha = 0.4$, making them far more robust to random pattern distortion, even though training is conducted on noise-free patterns (\cref{fig:memory} C). Interestingly, infomorphic networks cannot reliably encode very few patterns, as in this case many neurons receive the exact same input for every pattern, resulting in 0 bits of information in the receptive inputs.

\paragraph{Information Dynamics} We find an expected increase in redundant information $I_{\red}$ over the course of training (\cref{fig:supervised} B). Surprisingly, this increase is also present in networks that are seemingly above capacity, but then coincides with \emph{misinformative} unique contextual information $I_{\unq, C}<0$, indicating that each neuron's activity is not fully predictable by the other neurons in this case (Appendix \cref{fig:memory-appendix}).

\section{Discussion}
\label{sec:discussion}

In this work, we defined the infomorphic neuron, a new artificial neuron with two input classes and a flexible, parametrized local goal function derived from Partial Information Decomposition~\citep{williams2010nonnegative}. Like classical information theory, PID provides an abstract, high-level description of neural functioning, yet enriches this description with the additional structure of redundancy, uniqueness and synergy. This structure is inherited by the infomorphic neuron and leads to a highly interpretable and flexible description of local goals, independent of task, substrate, type of signals and encoding of information therein~\citep{wibral2015bits,wibral2017goal-function}.

The experiments conducted provide a proof of principle that the level of abstraction gained by an information-based approach does not compromise the ability of model neural networks to learn and solve diverse tasks.
Concretely, we find that maximizing the encoding of redundant information between the input and the label enables a single-layer network of infomorphic neurons to do supervised learning. Furthermore, we show that input information can be distributed between multiple infomorphic neurons in a recurrent network in an unsupervised learning task, by making each neuron maximize its encoding of unique input information with respect to the activity of other neurons. Finally, we find that maximizing the encoding of redundant information between an external input and the activity of other neurons in a recurrent infomorphic network leads to the formation of robust associative memories that exceed the memory capacity of classical Hopfield networks.

Notably, these experiments only explore a fraction of the available space of parameterized goal functions, and other terms of the goal function might become relevant in other learning scenarios. To demonstrate this, we construct primitive error neurons in the spirit of predictive coding (\cite{rao1999predictive, millidge2022predictive}; \cref{apx:error-neurons} and Appendix \cref{fig:error}), which receive simulated observations as receptive input and simulated model predictions as contextual input. They are trained to maximize synergistic information and both unique information atoms, while simultaneously minimizing redundant information. In a similar spirit, goals can be linearly combined to create more complex goal functions in other scenarios. One example of this are the weak incentives for unique and synergistic atoms in the supervised and associative memory experiments. We expect the residual entropy $H_{\res}$ to be another useful incentive to either reduce or artificially inject noise into infomorphic networks to improve learning.

Note that changing the hyper-parameters $\mbf \Gamma$ and the activation functions allows us to arbitrate between three very different learning tasks with little effort, providing practical evidence that our goal function $G$ is highly interpretable and provides an intuitively accessible understanding of the \emph{local goal of each neuron} in solving various tasks. Such interpretability is hard to establish in conventional ANNs, where a global error minimization goal is automatically back-propagated to the local level to adjust neuron parameters (e.g., \cite{lecun2015deep,samek2017explainable}). Furthermore, a similar understanding of local goal functions might be an insightful target in our description of biological neural networks, and ultimately help to bridge the gap between artificial and biological intelligent systems. To this end, the synthetic methodology of infomorphic networks can easily be combined with post-hoc analyses of trained BNNs and ANNs. In particular, it remains an open question which local information quantities these existing networks are effectively maximizing~\citep{wibral2015bits, ehrlich2023a, schulz2021gaba}.

\subsection{Future work}

Both the supervised and unsupervised experiments reported here focused on small single-layer neural networks, yet the ultimate strength of neural networks lies in their scalability to multilayered networks with a large number of neurons~\citep{montesinos2022fundamentals}. Currently, this scalability is not present in infomorphic networks due to a conundrum that is implicitly solved in backpropagation: In most architectures, a neuron does not only need to get a feedforward input and a context signal that conveys target information (like a reward, supervision or self-supervision signal), but additionally it requires knowledge about what other neurons in the same layer are coding for, such that it can choose to provide a complementary contribution with respect to the target~\citep{sacramento2018dendritic}. In backpropagation, this information flows to the neuron implicitly through the gradient signal from higher layers~\citep{whittington2019theories,sacramento2018dendritic}, yet in infomorphic networks it needs to be provided explicitly, because it fulfills a different role from the feed-back information: While the neuron typically needs to follow the feed-back signal (redundancy), it simultaneously needs to be different from the lateral signal (uniqueness). In follow-up work, we define an infomorphic neuron with three input classes that combines these ideas, leading to greatly improved supervised learning performance~\citep{schneider2025what}. 

Additionally, it has been shown that encoding for synergy plays a role in integration of information from multiple sources or sensory streams, both in the brain~\citep{luppi2020synergistic,luppi2022synergistic} and in ANNs~\citep{proca2024synergistic}. This could be tested constructively in infomorphic neurons with four input classes, which could be trained to extract information that is synergistic between two different receptive inputs, redundant with a contextual input and unique with respect to other lateral neurons.

Infomorphic networks also offer a natural connection to other information-based learning algorithms. Prominently, under the infomax principle~\citep{linsker1988self} it has been shown that in the low-noise regime neurons maximize global information encoding by finding unique, independent features, which is in line with uniqueness maximization in our unsupervised learning experiments. However, under high noise, as might be prevalent in biological networks~\citep{london2010sensitivity}, more cooperative, redundant representations emerge~\citep{linsker1988self}. It remains an intriguing open question whether increasing admixtures of redundancy to the individual neural goals of infomorphic neurons may lead them to develop similar noise-robust representations. To test this hypothesis, our current framework allows increasing the noise by introducing an additional $H_{\res}$ term or stronger weight decay. However, learning under constant noise is a constrained optimization problem and we leave the introduction of the required Lagrange multipliers to future work.

Despite their fully local computation, infomorphic neurons currently lack biological plausibility due to the complexity of the gradient equations (\cref{eq:G_wR,eq:G_wC}) as well as the memory-expensive histogram estimation method in \cref{sec:infomorphic_neuron}. Whether the gradients can be effectively approximated by simpler equations remains an open question for future work. However, a parameterized estimation of $p(R,C)$ combined with PID-estimators for mixed discrete-continuous variables would significantly reduce memory cost, e.g.\ to only 5 parameters for a two-dimensional Gaussian~\citep{kay2011coherent, kay1994information}, while for some tasks like supervised classification, more expressive multimodal distributions might be necessary. In addition, infomorphic neurons are functional, not anatomical, units and might thus be modeled by small microcircuits instead of individual spiking neurons.

Finally, notice that the high degree of interpretability of PID and our simple task setups allowed for a very intuitive reasoning about hyper-parameters with only minor fine tuning. However, more complex tasks with bigger infomorphic networks might require more systematic hyper-parameter optimization techniques, a variety of which are easily accessible in modern-day machine learning tools~\citep{bischl2023hyperparameter}. Fortunately, the resulting hyper-parameter sets will then still be formulated in the language of Partial Information Decomposition and thus potentially provide crucial insights into the optimal local goals to enable the self-organization and collaboration of local units to solve a variety of global tasks~\citep{schneider2025what}.


\subsection{Conclusion}
Leveraging Partial Information Decomposition, this work establishes the infomorphic neuron, a novel neuron model permitting the flexible and direct optimization of interpretable information-theoretic goals. Through several lines of experimentation, the versatiliy of these neurons to solve a variety of machine learning tasks has been demonstrated. We propose infomorphic neurons as abstract neuron models that can provide a new foundation for studying information processing in neural networks \emph{in the language of local goals}, opening up many exciting avenues for future research.

\section{Materials and Methods}
Here we provide the material and methods that apply to all our simulation experiments, while specific variations are discussed in the respective subsections of \cref{sec:experiments}.
In all simulation experiments, we run discrete-time neural networks. After choosing the network topology, the goal function and activation function of each neuron, we initialize all weights at small values. Stimuli vary by experiment and are always presented to the networks sequentially, without re-initialization of neural activities. For each experiment we segment the execution time into mini-batches of fixed length. After each mini-batch, we construct the full binned probability distribution $p_{Y, \tilde R, \tilde C}$ of each neuron from the histogram of its inputs and outputs, which are obtained by quantizing $R$ and $C$ in uniform bins. Assuming the histogram as constant, we compute the gradients of the information-theoretic neural goal function $G$ with respect to the weights according to \cref{eq:G_wR,eq:G_wC}, and conduct a weight update by applying gradient descent with a fixed learning rate.

To evaluate each experiment, we calculate performance metrics over the course of training by interleaving the training data stream with test data at regular intervals, while suppressing weight updates. These metrics include both the ``information dynamics'', i.e.~the size of all information atoms of each neuron, and more traditional performance metrics like task accuracy, which our algorithm does not explicitly optimize for. For details on the parameter choices and results of each experiment, we refer the reader to the respective experimental Sections \ref{sec:experiments}\ref{subsec:supervised},~\ref{subsec:unsupervised} and \ref{subsec:memory} as well as to \cref{apx:exps-params} and \cref{apx:exp-observ}.
The experiments have been implemented in Python and have been made available on GitLab at \url{https://gitlab.gwdg.de/wibral/infomorphic_networks}.

\subsection*{Data Availability}
The full derivation of the learning rules, parameters for all experiments, as well as supplementary figures providing additional information on the experiments are provided in the Appendix. Code for reproducing all experiments is available on GitLab at \url{https://gitlab.gwdg.de/wibral/infomorphic_networks}. The raw data of the experiments is accessible via G\"{o}ttingen Research Online Data at \url{https://doi.org/10.25625/0M1PTJ}. All other data are included in the manuscript and/or Appendix.

\paragraph{Acknowledgments.} 
We would like to thank William Phillips and Jim Kay for opening this line of research and fruitful discussions on Partial Information Decomposition and neural networks in general. We would like to thank Fabian Mikulasch, Matthias Loidolt, and Lucas Rudelt for extensive discussions on this topic. We would also like to thank Valentin Neuhaus, Kjartan van Driel, Paul Spitzner, Johannes Zierenberg, Aaron Gutknecht, Jonas Dehning, and the rest of the Wibral and Priesemann groups for their valuable comments and feedback. Finally, we would like to thank the reviewers for their insightful input. M.G. received a scholarship by the Champalimaud Foundation, the German Academic Scholarship Foundation, as well as the FIH-D Scholarship by the Department of Chemistry and Applied Biosciences at Eidgenössische Technische Hochschule Zürich. A.M. and M.W. are employed at the Göttingen Campus Institute for Dynamics of Biological Networks funded by the Volkswagenstiftung. M.W. was supported by the flagship science initiative of the European Commission’s Future and Emerging Technologies program under the Human Brain project, HBP-SP3.1-SGA1-T3.6.1.A.C.S. and V.P. acknowledge support from the Max Planck Society and the Deutsche Forschungsgemeinschaft (German Research Foundation) under Germany’s Excellence Strategy-EXC 2067/1-390729940 as well as the RTG 2906 Curiosity. V.P. and M.W. received funding from the Deutsche Forschungsgemeinschaft (German Research Foundation) via the SFB 1528 ``Cognition of Interaction'' - project-ID 454648639. D.A.E. and M.W. were supported by a funding from the Ministry for Science and Education of Lower Saxony and the Volkswagen Foundation through the ``Niedersächsisches Vorab'' under the program ``Big Data in den Lebenswissenschaften''-project ``Deep learning techniques for association studies of transcriptome and systems dynamics in tissue morphogenesis.'' Open access funding provided by the Max Planck Society.


\appendix
\section{Derivation of the learning rules\label{apx:gradients}}
In the following we provide a detailed analytical derivation of the learning rules.

The value of the goal function is dependent on the neuron-specific parameter vector $\mbf \Gamma$ and the joint distribution $p(R,C,Y)$ of a neuron's integrated receptive and contextual inputs and its output. Learning happens by changing the afferent weights $\mbf w_R$ and $\mbf w_C$ which determine how the inputs to the neuron influence its output. To learn these weights via gradient descent, we require the gradients $\frac{\partial G}{\partial\mbf w_R}$ and $\frac{\partial G}{\partial\mbf w_C}$. Starting from the definition of $G$, we first make use of the fact that a computable expression for one PID atom, in this case the redundant information $I_{\red}(Y: R,C):= I^{\sx}_\cap(Y: R;C)$, in combination with the rules of classical information theory suffices to quantify all four PID atoms~\citep{gutknecht2021bits}. Note that we stick to the notation ``;'' in $I^{\sx}_cap$ introduced by Makkeh et al.~\citep{makkeh2021isx} which stands for the union of $R$ and $C$ stressing that $I^{\sx}_\cap$ is based on union-exclusions. This allows us to re-parametrize
\begin{equation*}
    \begin{split}
        G(Y: R, C) &= \Gamma_{\unq, R}I_{\unq, R}(Y: R, C) + \Gamma_{\unq, C}I_{\unq, C}(Y: R, C) + \Gamma_{\red} I_{\red}(Y: R, C) + \Gamma_{\syn}I_{\syn}(Y: R, C) +\Gamma_{\res}H(Y\mid R, C)\\
        &=: \gamma_{Y}H(Y) + \gamma_{Y\mid R}H(Y\mid R) + \gamma_{Y\mid C} H(Y\mid  C) +\gamma_{Y\mid R,C}H(Y\mid R, C) +\gamma_{\red}I_{\red}(Y: R, C)\;,\\
    \end{split}
\end{equation*}

with the parameter transformation
\begin{equation*}
    \mbf \Gamma = 
    \begin{pmatrix}
    \Gamma_{\unq, R}\\
    \Gamma_{\unq, C}\\
    \Gamma_{\red}\\
    \Gamma_{\syn}\\
    \Gamma_{\res}
    \end{pmatrix} =: \begin{pmatrix}
                         1 &  0 &  1 &  0 &  0\\
                         1 &  1 &  0 &  0 &  0\\
                         1 &  0 &  0 &  0 &  1\\
                         1 &  1 &  1 &  0 &  0\\
                         1 &  1 &  1 &  1 &  0
            \end{pmatrix}\begin{pmatrix}
                            \gamma_{Y}\\
                            \gamma_{Y\mid R}\\
                            \gamma_{Y\mid C}\\
                            \gamma_{Y\mid R,C}\\
                            \gamma_{\red}
                        \end{pmatrix}\quad \iff \quad
    \mbf \gamma :=
    \begin{pmatrix}
    \gamma_{Y}\\
    \gamma_{Y\mid R}\\
    \gamma_{Y\mid C}\\
    \gamma_{Y\mid R,C}\\
    \gamma_{\red}
    \end{pmatrix} := \begin{pmatrix}
                         1 &  1 &  0 & -1 & 0\\
                        -1 &  0 &  0 &  1 & 0\\
                         0 & -1 &  0 &  1 & 0\\
                         0 &  0 &  0 & -1 & 1\\
                        -1 & -1 &  1 &  1 & 0
            \end{pmatrix}\begin{pmatrix}
                            \Gamma_{\unq, R}\\
                            \Gamma_{\unq, C}\\
                            \Gamma_{\red}\\
                            \Gamma_{\syn}\\
                            \Gamma_{\res}
                        \end{pmatrix}\;.
\end{equation*}

Next, notice that $G$ is only dependent on the weights through the conditional firing probability of the neuron $\theta(r,c) = P(Y=+1\mid R= r, C = c )$. Ultimately, we want to arrive at an expression of the form
\begin{equation*}
    \begin{split}
        \frac{\partial G}{\partial\mbf w_R} &= \left\langle\frac{\partial g(r,c)}{\partial\theta(r,c)}\frac{\partial\theta(r,c)}{\partial\mbf w_R}\right\rangle_{r,c}\\
        \frac{\partial G}{\partial\mbf w_C} &= \left\langle\frac{\partial g(r,c)}{\partial\theta(r,c)
        }\frac{\partial\theta(r,c)}{\partial\mbf w_C}\right\rangle_{r,c}\;,
    \end{split}
\end{equation*}
where we expressed $G$ as an expectation over the joint distribution of the integrated inputs $p(r,c)$. Introducing a quantity $g$ that we can express for every value of $r$ and $c$ allows us to then apply the chain rule, and is possible due to the fact that all information-theoretic terms in $G$ are point-wise measures, including the redundant information measure~\citep{makkeh2021isx}
\begin{equation*}
    I_{\red}(Y:R,C) = I^{\sx}_\cap(Y: R;C) = \left\langle i^{\sx}_\cap(y: r; c)\right\rangle_{r,c,y}
    := \left\langle\log\frac{p_\theta(y \mid r\cup c)}{p_\theta(y)}\right\rangle_{r,c,y}\;,
\end{equation*}
where we have explicitly marked $\theta$-dependence by a subscript, and introduced the term
\begin{equation*}
        p_\theta(y \mid r\cup c) = \frac{p_\theta(y, r\cup c)}{p(r\cup c)}
\end{equation*}
with
\begin{equation*}
    \begin{split}
        p(r\cup c) &= p(r)+p(c)-p(r,c)\\
        p_\theta(y, r\cup c) &= p(y,r)+p(y,c)-p(y,r,c)\\
        &=\left\langle\theta(r,c')p(r,c')\right\rangle_{c'}+\left\langle\theta(r',c)p(r',c)\right\rangle_{r'}-\theta(r,c)p(r,c).
    \end{split}
\end{equation*}
Inserting definitions and separating the terms for $y=1$ and $y=-1$, we can now write

\begin{alignat*}{6}
    G(Y: R, C)
    =&-\gamma_{Y} &&\bigl[\quad p(+1)&&\log p(+1) &&+ p(-1)&&\log p(-1)&&\bigr]\\
    &-\gamma_{Y\mid R} &&\bigl\langle\quad p(+1|r)&&\log p(+1|r) &&+ p(-1|r)&&\log p(-1|r)&&\bigr\rangle_r\\
    &-\gamma_{Y\mid C} &&\bigl\langle\quad p(+1|c)&&\log p(+1|c) &&+ p(-1|c)&&\log p(-1|c)&&\bigr\rangle_c\\
    &-\gamma_{Y\mid R,C} &&\bigl\langle\quad p(+1|r,c)&&\log p(+1|r,c) &&+ p(-1|r,c)&&\log p(-1|r,c)&&\bigr\rangle_{r,c}\\
    &-\gamma_{\red} &&\bigl[\quad p(+1)&&\log p(+1) &&+ p(-1)&&\log p(-1)&&\bigr]\\
    &+\gamma_{\red} &&\bigl\langle\quad p(+1|r,c)&&\log p(+1|r\cup c) &&+ p(-1|r,c)&&\log p(-1|r\cup c)&&\bigr\rangle_{r,c}\\
    =&-\gamma_{Y} &&\bigl\langle\quad \theta&&\log \langle \theta\rangle_{r,c} &&+ (1-\theta)&&\log (1-\langle \theta\rangle_{r,c})&&\bigr\rangle_{r,c}\\
    &-\gamma_{Y\mid R} &&\bigl\langle\quad \theta&&\log \langle \theta\rangle_{c|r} &&+ (1-\theta)&&\log (1-\langle \theta\rangle_{c|r})&&\bigr\rangle_{r,c}\\
    &-\gamma_{Y\mid C} &&\bigl\langle\quad \theta&&\log \langle \theta\rangle_{r|c} &&+ (1-\theta)&&\log (1-\langle \theta\rangle_{r|c})&&\bigr\rangle_{r,c}\\
    &-\gamma_{Y\mid R,C} &&\bigl\langle\quad \theta&&\log \theta &&+ (1-\theta)&&\log (1-\theta)&&\bigr\rangle_{r,c}\\
    &-\gamma_{\red} &&\bigl\langle\quad \theta&&\log \langle \theta\rangle_{r,c} &&+ (1-\theta)&&\log (1-\langle \theta\rangle_{r,c})&&\bigr\rangle_{r,c}\\
    &+\gamma_{\red} &&\bigl\langle\quad \theta&&\log p_\theta(+1|r\cup c) &&+ (1-\theta)&&\log (1-p_\theta(+1|r\cup c))&&\bigr\rangle_{r,c}\;,
\end{alignat*}
where in the second equation we have made the dependence on $\theta$ explicit and written all terms in the same expectation value. From the first equation, it is straightforward to see that differentiating the logarithms leads to terms that cancel in all cases except the last line of the redundant information. Taking into account that
\begin{equation*}
    \begin{split}
        \theta(r,c) = \sigma(A(r, c)) &= \sigma(A(\mbf w_R^T\mbf x_R, \mbf w_C^T \mbf x_C))\\
        \Rightarrow\frac{\partial\theta(r,c)}{\partial\mbf w_R} &= \theta(r,c)(1-\theta(r,c))\frac{\partial A}{\partial r}\mbf x_R\\
        \frac{\partial\theta(r,c)}{\partial\mbf w_C} &=\theta(r,c)(1-\theta(r,c))\frac{\partial A}{\partial c}\mbf x_C\;,
    \end{split}
\end{equation*}
with the sigmoid function $\sigma(\xi)=\frac{1}{1+e^{-\xi}}; \frac{d\sigma}{d\xi}=\sigma(\xi)(1-\sigma(\xi))$, we thus obtain

\begin{equation*}\label{eq:learning-rules}
    \begin{split}
        \frac{\partial G}{\partial\mbf w_R} &= \left\langle \frac{\partial g(r,c)}{\partial\theta(r,c)}\theta(r,c)(1-\theta(r,c))\frac{\partial A}{\partial r}\mbf x_R\right\rangle_{r,c}\\
        \frac{\partial G}{\partial\mbf w_C}  &= \left\langle\frac{\partial g(r,c)}{\partial\theta(r,c)}\theta(r,c)(1-\theta(r,c))\frac{\partial A}{\partial c}\mbf x_C\right\rangle_{r,c}
    \end{split}
\end{equation*}
where
\begin{equation*}
    \begin{split}
        \frac{\partial g(r,c)}{\partial\theta(r,c)} &=-(\gamma_{Y}+\gamma_{\red}) \log\frac{\langle \theta\rangle_{r,c}}{1-\langle \theta\rangle_{r,c}} -\gamma_{Y\mid R} \log\frac{\langle \theta\rangle_{c|r}}{1-\langle \theta\rangle_{c|r}} -\gamma_{Y\mid C} \log\frac{\langle \theta\rangle_{r|c}}{1-\langle \theta\rangle_{r|c}} +\gamma_{Y\mid R,C} A(r,c)\\
        &+\gamma_{\red} \left[\log\left(\frac{p_\theta(+1|r\cup c)}{1-p_\theta(+1|r\cup c)}\right) + \left(\frac{\theta}{ p_\theta(+1|r\cup c)} - \frac{1-\theta}{1- p_\theta(+1|r\cup c)}\right) p(r\cup c)\right].
    \end{split}
\end{equation*}
Taking the expectation values over $p(r,c)$ instead of $p(\mbf x_R, \mbf x_C)$ here does not change the outcome when working with the empirically sampled inputs, but significantly simplifies notation and calculations \citep{kay1994information}. Note that $\frac{\partial G}{\partial\mbf w_R}$ and $\frac{\partial G}{\partial\mbf w_C}$ inherit the term $\theta(r,c)(1-\theta(r,c))$ from the derivative of the sigmoid function. This term is small whenever the firing probability is far from $\theta(r,c)=\frac{1}{2}$, and gives inputs that lead to highly stochastic firing increased influence on learning.

To arrive at the empirical gradients, we substitute the expectation value over the joint distribution of the integrated inputs by an empirical average over the $t$-th batch and obtain
\begin{equation*}
    \begin{split}
        \frac{\partial \hat{G}^t}{\partial\mbf w_R} &= \frac{1}{\lvert B^t\rvert} \sum_{\{\mbf x_R, \mbf x_C\}\in B^t} \left[\left.\frac{\partial g}{\partial\theta}\right\rvert_{\tilde r, \tilde c}\theta(\tilde r,\tilde c)(1-\theta(\tilde r,\tilde c))\left.\frac{\partial A}{\partial r}\right\rvert_{\tilde r, \tilde c}\mbf x_R\right]\\
        &:=\frac{1}{\lvert B^t\rvert} \sum_{\{\mbf x_R, \mbf x_C\}\in B^t} \left[f_{p(R,C)}^{\mbf\Gamma}(\tilde r, \tilde c)\left.\frac{\partial A}{\partial r}\right\rvert_{\tilde r, \tilde c}\mbf x_R\right]\\
        \frac{\partial \hat{G}^t}{\partial\mbf w_C}  &= \frac{1}{\lvert B^t\rvert} \sum_{\{\mbf x_R, \mbf x_C\}\in B^t} \left[\left.\frac{\partial g}{\partial\theta}\right\rvert_{\tilde r, \tilde c}\theta(\tilde r,\tilde c)(1-\theta(\tilde r,\tilde c))\left.\frac{\partial A}{\partial c}\right\rvert_{\tilde r, \tilde c}\mbf x_C\right]\\
        &:= \frac{1}{\lvert B^t\rvert} \sum_{\{\mbf x_R, \mbf x_C\}\in B^t} \left[f_{p(R,C)}^{\mbf\Gamma}(\tilde r, \tilde c)\left.\frac{\partial A}{\partial c}\right\rvert_{\tilde r, \tilde c}\mbf x_C\right]\;.
    \end{split}
\end{equation*}

where $B^t$ denotes the batch corresponding to training time $t$, and the batch size $\lvert B^t\rvert = N_\mathrm{tr}$ or $\lvert B^t\rvert = N_\mathrm{te}$ is constant in all our experiments. Furthermore, $(\tilde r, \tilde c)$ are the binned observations, and we use $h(\tilde r, \tilde c)$ as a shorthand to indicate that we are evaluating any function $h$ at the bin center of the bin corresponding to the tuple $(r,c)$. The notation with $f_{p(R,C)}^{\mbf\Gamma}(\tilde r, \tilde c)$ corresponds to equations \eqref{eq:G_wR} and \eqref{eq:G_wC}.

Finally, this leads to the update rules
\begin{equation*}
    \begin{split}
        \mbf w_R^{t+1} &= \mbf w_R^t + \eta \frac{\partial \hat{G}^t}{\partial\mbf w_R}\\
        \mbf w_C^{t+1} &= \mbf w_C^t + \eta \frac{\partial \hat{G}^t}{\partial\mbf w_C}\;.
    \end{split}
\end{equation*}

\section{Experiments: parameters and statistics\label{apx:exps-params}}
In the following we will give a full account of the setup and parameter choices of the experiments that will allow reproduction of our results. We will first explain each parameter and then tabulate the parameter values for all three experimental schemes (supervised, unsupervised and memory).

\subsection{Explanation of parameters}
Each infomorphic neuron has a set of parameters that determine its function and goal. Additionally, in each task, the training of these neurons might slightly vary. The training and individual neuron parameters are: 
\begin{itemize}
    \item \textbf{Training:}
    \begin{itemize}
        \item phases: In the unsupervised learning experiment, neurons undergo training in two phases, each with different parameter settings. We provide the number of training steps for each phase as a tuple of integers in the ``phases'' parameter. In the other experiments, this parameter is simply an integer indicating the number of training steps.    
        \item $N_\mathrm{tr}$ and $N_\mathrm{te}$: The number of training and testing samples in a mini-batch per training step, respectively.
        \item $m_{\mathrm{rep}}$: The number of times an individual input is presented. We introduce this parameter as the unsupervised and memory experiments require presentation of each input over multiple time steps. 
    \end{itemize}
    \item \textbf{Learning:}
    \begin{itemize}
        \item $b_\mathrm{init}$: The order of magnitude of the initialized weights. For example $b_\mathrm{init}=0.1$ means that the weights are initialized uniformly at random in the interval $[-0.1, 0.1]$. 
        \item goal function parameters $\Gamma$: The parameters defining which information contributions a neuron maximizes or minimizes. We use the vector notation $\mbf \Gamma = (\Gamma_{\unq, R}, \Gamma_{\unq, C}, \Gamma_{\red}, \Gamma_{\syn}, \Gamma_{\res})$. For example, in the supervised learning experiment, $\Gamma=(0.1,0.1,1,0.1,0)$. The parameter vector $\mbf\Gamma$ does not need to be of unit length and deviations from unit length will effectively act as a scaling factor on the learning rate.
        \item learning rate $\eta$: The learning rate for the gradient ascent utilized to maximize the goal function.
        \item pullback rate $\lambda$: The weights decay factor. In the first training phase of the unsupervised learning experiment, the receptive weights are pulled back by $\lambda$ at each step to keep them from growing. This pullback acts as follows: $w_R= w_R - 2\lambda w_R + \eta\nabla(w_R)$ where $\nabla(w_R)$ is the gradient update at this step. 
    \end{itemize}
    \item \textbf{Input Integration:}
    \begin{itemize}
        \item $n_{\mathrm{receptive}}$: The dimension of the input vector $\mbf X_R$, i.e. the number of receptive inputs. Evidently, $n_{\mathrm{receptive}}$ is also the size of the vector $\mbf w_R$.
        \item $n_{\mathrm{contextual}}$: The dimension of the input vector $\mbf X_C$, i.e. the number of contextual inputs. Evidently, $n_{\mathrm{contextual}}$ is also the size of the vector $\mbf w_C$.
        \item $J_R$ and $J_C$: The interval in which the $R$ and $C$ values are binned, respectively. Any realizations $r$ or $c$ exceeding the limits of their respective interval are aggregated into two additional bins of infinite width at either side.
        \item $n_{\mathrm{bins},r}$ and $n_{\mathrm{bins,}c}$: The number of equal-width bins in the intervals $J_R$ and $J_C$, respectively. The total number of bins is $(n_{\mathrm{bins},r} + 2) \times (n_{\mathrm{bins,}c} + 2)$ due to the bins of infinite width bounding the respective intervals. The discrete random variables resulting from the binning of $R$ and $C$ are denoted as $\tilde R$ and $\tilde C$. 
    \end{itemize}
\end{itemize}
\subsection{Supervised learning\label{apx:params-supervised}}
In this experiment 100 networks were run. All networks were identical in architecture, with a single layer of 10 neurons, but initialized at different sets of random weights. During training, the 28-by-28 MNIST pixel images were presented in random order as the $784$-dimensional vector $\mbf X_R$, with each pixel value rescaled to the interval $[-1,1]$. The label was presented as a one-hot representation $\mbf X_C$. During testing, the network received the MNIST image as in training, however, a constant vector of zeroes on $\mbf X_C$. Note that during training $X_C \in {-1,1}$, so a value of $X_C = 0$ is equidistant from both possible values during training. The training and individual neuron parameters are summarized in \cref{tab:sup}.

\begin{table}[!ht]
    \caption{\label{tab:sup}\textit{Parameters of the infomorphic neurons in the supervised learning experimental scheme}.}
    \centering
    \begin{tabular}{|c|c||c|}
        \toprule
        \multicolumn{3}{|c|}{Training}\\
        \cmidrule{1-3}
        \multicolumn{1}{|c|}{Parameter}    &\multicolumn{1}{c||}{Value}    &\multicolumn{1}{c|}{Comment}\\
            \cmidrule(r){1-2}\cmidrule(lr){3-3}
            Phases & 800 & a single phase of training with 800 training steps (batches)\\
            $N_\mathrm{tr}$ & 1000& batch sampled uniformly from the 60000 MNIST training images\\
            $N_\mathrm{te}$ & 1000& batch sampled uniformly from the 10000 MNIST testing images\\
            $m_{\mathrm{rep}}$ & 1 & no repetition is needed\\
            \cmidrule{1-3}\morecmidrules\cmidrule{1-3}
        \multicolumn{3}{|c|}{Learning}\\ 
        \cmidrule{1-3}
        \multicolumn{1}{|c|}{Parameter}    &\multicolumn{1}{c||}{Value}    &\multicolumn{1}{c|}{Comment}\\
            \cmidrule(r){1-2}\cmidrule(lr){3-3}
            $b_\mathrm{init}$ & 0.01 & scale of initialization\\
            $\mbf \Gamma$ &(0.1, 0.1, 1, 0.1, 0) & goal parameters ($\Gamma_{\unq, R}$, $\Gamma_{\unq, C}$, $\Gamma_{\red}$, $\Gamma_{\syn}$, $\Gamma_{\res}$)\\
            $\eta$ & 1.0 & learning rate\\
            $\lambda$ & 0.0 & no pullback needed\\
            \cmidrule{1-3}\morecmidrules\cmidrule{1-3}
        \multicolumn{3}{|c|}{Input Integration}\\ 
        \cmidrule{1-3}
        \multicolumn{1}{|c|}{Parameter}    &\multicolumn{1}{c||}{Value}    &\multicolumn{1}{c|}{Comment}\\
            \cmidrule(r){1-2}\cmidrule(lr){3-3}
            $n_{\mathrm{receptive}}$ & 784 & MNIST image pixel size\\
            $n_{\mathrm{contextual}}$ & 1& one element of a one-hot label vector\\
            $J_R$ & $[-20, 20]$ & a smaller range might hinder the learning\\
            $J_C$ & $[-20, 20]$ & a smaller range might hinder the learning\\
            $n_R$-bins & 200 & uniform bin-size is 0.1\\
            $n_C$-bins & 200 & uniform bin-size is 0.1\\
            \bottomrule
        \end{tabular}
    \end{table}

\subsection{Unsupervised Learning\label{apx:params-unsupervised}}
In this experiment 300 networks were run. All networks were identical in architecture with a single layer of 8 neurons, but initialized at different sets of random weights. The neurons were recurrently connected via their contextual inputs $\mbf x_C$. During training and testing, an 8-by-8 grid of pixels was received as the $64$-dimensional vector $\mbf X_R$ with discrete pixel values in $\{-1,1\}$. Training was split into two phases: the first with a weight decay to keep the weights low and allow communication between the neurons, while in the second phase the weight decay was turned off to make the magnitude of all weights increase to the final solution. The training and individual neuron parameters are summarized in \cref{tab:unsup}.

\begin{table}[!ht]
    \caption{\label{tab:unsup}\textit{Parameters of the infomorphic neurons in the unsupervised learning experimental scheme}.} 
    \centering
    \begin{tabular}{|c|c||c|}
        \toprule
        \multicolumn{3}{|c|}{Training}\\
        \cmidrule{1-3}
        \multicolumn{1}{|c|}{Parameter}    &\multicolumn{1}{c||}{Value}    &\multicolumn{1}{c|}{Comment}\\
            \cmidrule(r){1-2}\cmidrule(lr){3-3}
            Phases & $(50,50)$ & two phases of training with 50 training steps (batches) each, first with weight decay activated\\
            $N_\mathrm{tr}$ & 1000& batch generated randomly from a distribution of independent bars with $p=0.5$ for each bar\\
            $N_\mathrm{te}$ & 1000& batch generated randomly from a distribution of independent bars with $p=0.5$ for each bar\\
            $m_{\mathrm{rep}}$ & 8 & each sample is presented 8 consecutive time steps\\
            \cmidrule{1-3}\morecmidrules\cmidrule{1-3}
        \multicolumn{3}{|c|}{Learning}\\ 
        \cmidrule{1-3}
        \multicolumn{1}{|c|}{Parameter}    &\multicolumn{1}{c||}{Value}    &\multicolumn{1}{c|}{Comment}\\
            \cmidrule(r){1-2}\cmidrule(lr){3-3}
            $b_\mathrm{init}$ & 0.1 & scale of initialization\\
            $\mbf \Gamma$ &(1, 0, 0, 0, 0) & goal parameters ($\Gamma_{\unq, R}$, $\Gamma_{\unq, C}$, $\Gamma_{\red}$, $\Gamma_{\syn}$, $\Gamma_{\res}$), equal in both learning phases\\
            $\eta$ & (10.0, 1.0)& higher learning rate during weight decay phase \\
            $\lambda$ & (0.28, 0.0) & initial phase of weight decay \\
            \cmidrule{1-3}\morecmidrules\cmidrule{1-3}
        \multicolumn{3}{|c|}{Input Integration}\\ 
        \cmidrule{1-3}
        \multicolumn{1}{|c|}{Parameter}    &\multicolumn{1}{c||}{Value}    &\multicolumn{1}{c|}{Comment}\\
            \cmidrule(r){1-2}\cmidrule(lr){3-3}
            $n_{\mathrm{receptive}}$ & 64 & size of the input\\
            $n_{\mathrm{contextual}}$ & 7& one per each neuron with no self connections\\
            $J_R$ & $[-25, 25]$ & a smaller range might hinder the learning\\
            $J_C$ & $[-25, 25]$ & a smaller range might hinder the learning\\
            $n_R$-bins & 500 & uniform bin-size is 0.05; larger size might hinder the learning\\
            $n_C$-bins & 500 & uniform bin-size is 0.05; larger size might hinder the learning\\
            \bottomrule
        \end{tabular}
    \end{table}
    
\subsection{Associative Memory\label{apx:params-memory}}
In this experiment 425 individual networks of 100 neurons were run. The networks are clustered into 17 groups of 25 networks each; each of which was trained on a different number of memory patterns in $\{1,2, 4, 6, 8, 12, 15, 20, 25, 30,\allowbreak  35, 40, 45, 50, 55, 60\}$. All memory patterns were chosen as random $100$-dimensional binary vectors with exactly $50$ entries being $+1$ and $50$ entries being $-1$, indicated as $(p=0.5)$-sparsity in the main text. All networks were identical in architecture with a single layer of 100 neurons, but initialized at different sets of random weights. The neurons were recurrently connected via their contextual weights $\mbf w_C$, excluding self-connections. During training, each neuron received a single input $x_R \in \{-1,1\}$ at every time step, i.e. one bit of the presented memory pattern. During testing (recall), the pattern was presented only in the first time step, then for later steps $x_R$ was set to  0 for all neurons. For each of the 19 consecutive time steps of the recall, the responses of the neurons were read out and compared with the originally presented pattern using cosine similarity. In the main text we report the similarity in the last recall time step as the accuracy. The training and individual neuron parameters are summarized in \cref{tab:memo}.  

\begin{table}[!ht]
    \caption{\label{tab:memo}\textit{Parameters of the infomorphic neurons in the associative memory experimental scheme}.}
    \centering
    \begin{tabular}{|c|c||c|}
        \toprule
        \multicolumn{3}{|c|}{Training}\\
        \cmidrule{1-3}
        \multicolumn{1}{|c|}{Parameter}    &\multicolumn{1}{c||}{Value}    &\multicolumn{1}{c|}{Comment}\\
            \cmidrule(r){1-2}\cmidrule(lr){3-3}
            Phases & 300 & a single phase of training with 300 epochs\\
            $N_\mathrm{tr}$ & 200& sampled from the set of input patterns\\
            $N_\mathrm{te}$ & 200& sampled from the set of input patterns\\
            $m_{\mathrm{rep}}$ & 8 & each sample is presented 8 consecutive times\\
            \cmidrule{1-3}\morecmidrules\cmidrule{1-3}
        \multicolumn{3}{|c|}{Learning}\\ 
        \cmidrule{1-3}
        \multicolumn{1}{|c|}{Parameter}    &\multicolumn{1}{c||}{Value}    &\multicolumn{1}{c|}{Comment}\\
            \cmidrule(r){1-2}\cmidrule(lr){3-3}
            $b_\mathrm{init}$ & 0.1 & scale of initialization\\
            $\mbf \Gamma$ &(0.1, 0.1, 1, 0.1, 0) & goal parameters ($\Gamma_{\unq, R}$, $\Gamma_{\unq, C}$, $\Gamma_{\red}$, $\Gamma_{\syn}$, $\Gamma_{\res}$)\\
            $\eta$ & 0.48 & $\eta=0.5$ does not change the result \\
            $\lambda$ & 0.0 & no weight pullback\\
            \cmidrule{1-3}\morecmidrules\cmidrule{1-3}
        \multicolumn{3}{|c|}{Input Integration}\\ 
        \cmidrule{1-3}
        \multicolumn{1}{|c|}{Parameter}    &\multicolumn{1}{c||}{Value}    &\multicolumn{1}{c|}{Comment}\\
            \cmidrule(r){1-2}\cmidrule(lr){3-3}
            $n_{\mathrm{receptive}}$ & 1 & bit of the presented pattern\\
            $n_{\mathrm{contextual}}$ & 99& a one per each neuron with no self connections\\
            $J_R$ & $[-20, 20]$ & a smaller range might hinder the learning\\
            $J_C$ & $[-20, 20]$ & a smaller range might hinder the learning\\
            $n_R$-bins & 20 & uniform bin-size is 2; smaller size doesn't affect the learning\\
            $n_C$-bins & 20 & uniform bin-size is 2; smaller size doesn't affect the learning\\
            \bottomrule
        \end{tabular}
    \end{table}

\section{Experiments: Supplementary Observables\label{apx:exp-observ}}
In this section we show observables that complement the main observables shown for each experiment in the main paper. These observables are meant to give a more in depth account for various performance measures of the infomorphic networks in the supervised, unsupervised, and memory tasks.

\subsection{Supervised learning} 
\begin{figure}[!ht]
    \centering
    \includegraphics[width=\textwidth]{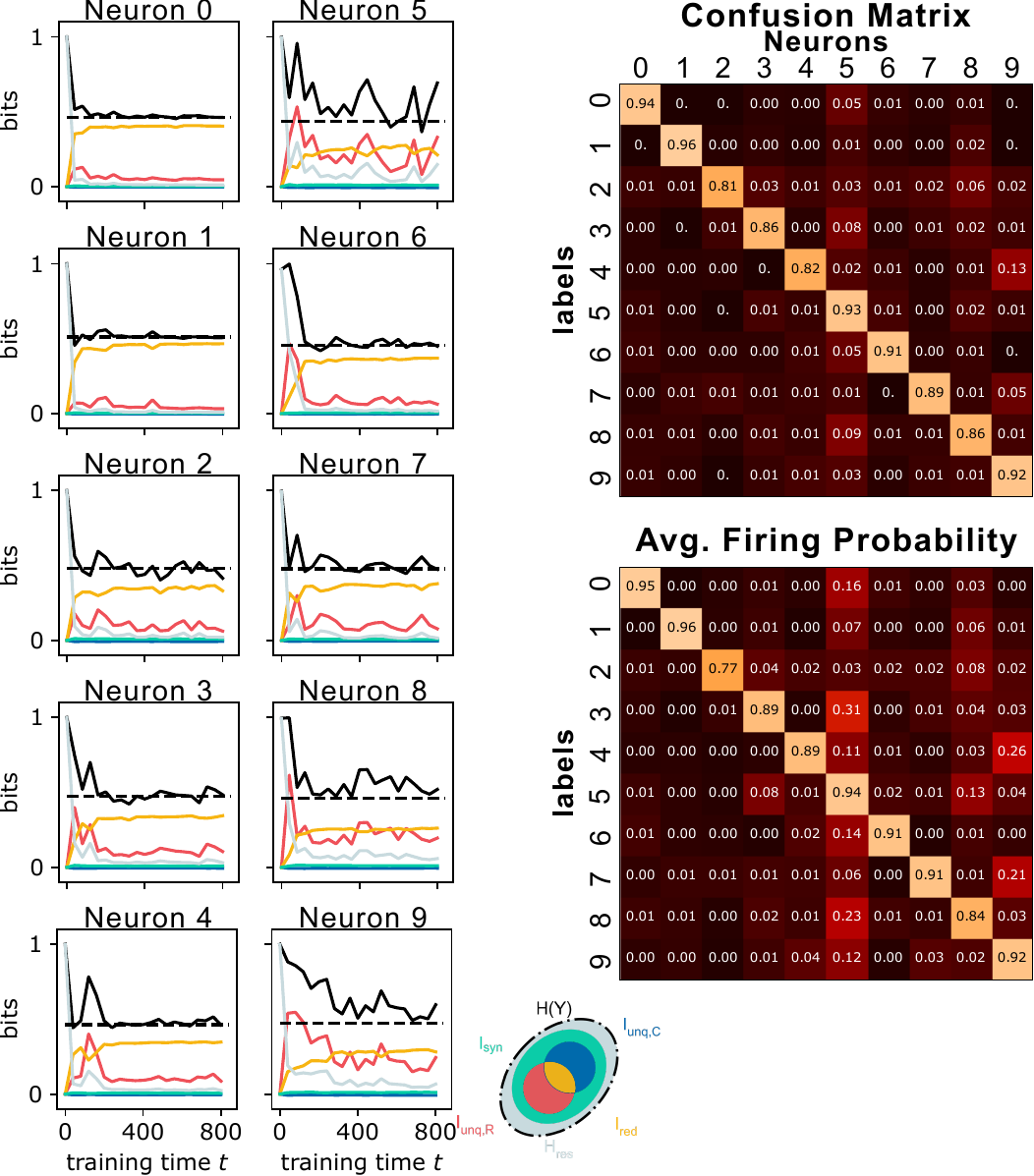}
    \caption{\textbf{In supervised learning, high redundant information $I_{\red}$ and low unique receptive information $I_{\unq,R}$ are associated with high classification accuracy.} \textbf{(Left)} Time evolution of information quantities for each neuron in a single, randomly chosen network performing supervised learning. Neuron $i$ corresponds to label $i$. The dashed line shows the empirical entropy of the binary one-vs-all distribution of each label in the test data set. \textbf{(Top right)} The confusion matrix for this network, after applying a winner-take-all readout. \textbf{(Bottom right)} The firing probability of each neuron for each digit, averaged over test images.
    \label{fig:mnist-pid}}
\end{figure}

\cref{fig:mnist-pid} shows the local information contributions of all ten neurons from one randomly chosen network, the respective confusion matrix and the averaged firing probability of each neuron for each label. High sensitivity and specificity of a neuron to its assigned digit correspond to high redundant information $I_{\red}$ and low values of all other information contributions (e.g. neurons 0, 1, and 6). In particular, in these cases both the entropy of the neuron and its redundant information approach the entropy of the respective label in the test data set. High firing probability of a neuron for digits from the wrong class (false positives) corresponds to a reduction of its redundant information $I_{\red}$ and an increase of the unique information of its receptive inputs $I_{\unq, R}$ (e.g. neurons 5, 8, and 9). Lower firing probability for the correct digit (fewer true positives) without high firing probability for wrong digits (few false positives) results in a weaker decrease of redundant and increase of unique information (e.g. neurons 2, 3, 4, 7).

To compare learning outcome and accuracy of the infomorphic networks, we additionally trained a logistic regression model on one-vs-all classification of MNIST digits for 3000 iterations with vanilla gradient descent and a learning rate of $\eta = 1$ (adapted from \citep{Mohapatra_2020}).

\begin{figure}[!ht]
    \centering
    \includegraphics[width=\textwidth]{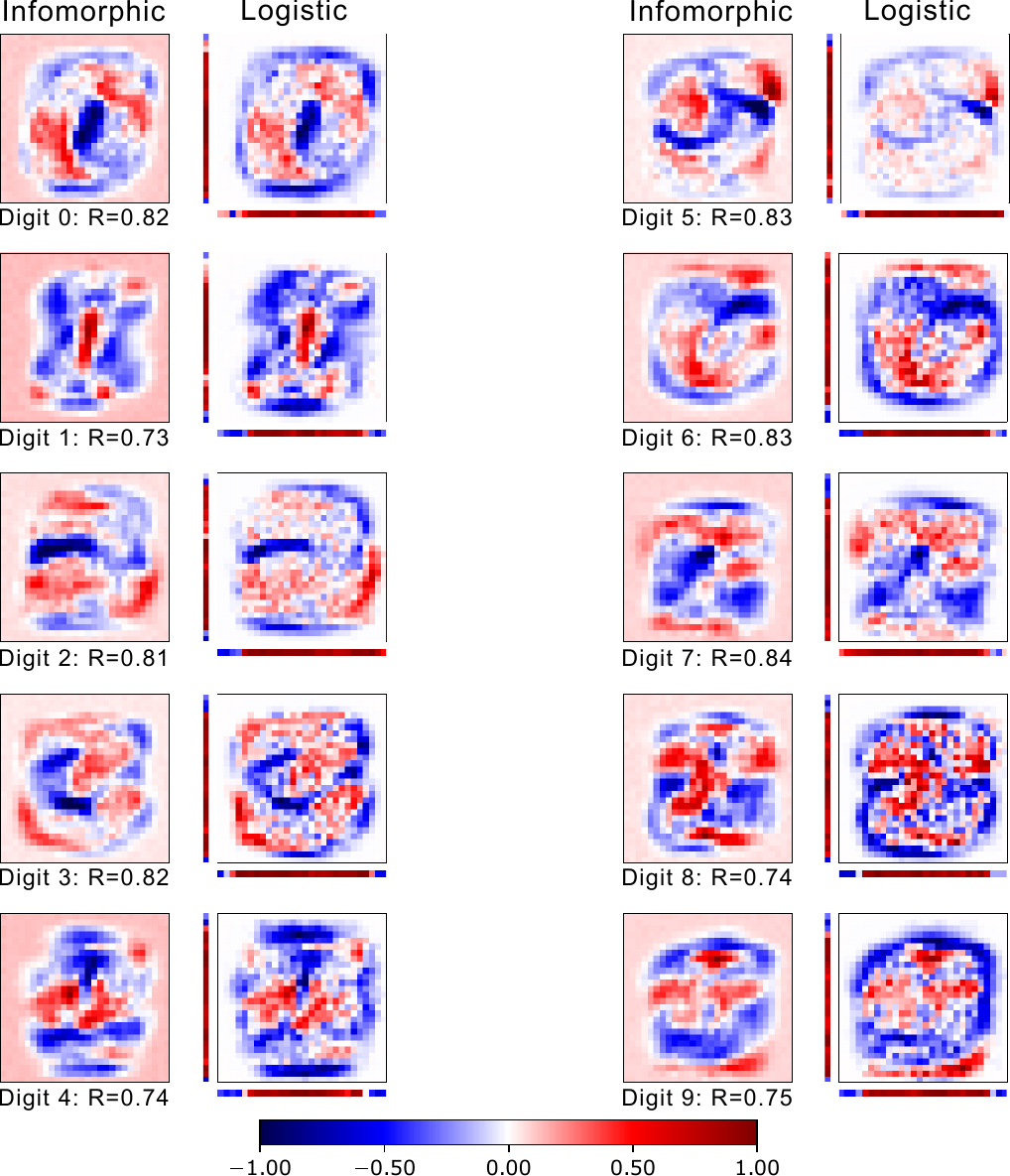}
    \caption{
   \textbf{The receptive fields of supervised infomorphic learning are similar to those of one-vs-all logistic regression.} First and third column: Receptive fields of all neurons after training in a randomly chosen supervised infomorphic network. Second and fourth column: Corresponding receptive fields obtained from one-vs-all logistic regression with vanilla gradient descent. The depicted receptive fields are centered at a weight of $w_R = 0$ and re-scaled to the interval $[-1,1]$. $R$-values indicate cosine similarity between corresponding receptive fields. The colored vectors bordering the receptive fields of logistic regression indicate row-by-row and column-by-column cosine similarity between corresponding receptive fields. Note that all values are on the same scale, indicated by the color bar at the bottom.}
    \label{fig:mnist-receptive-fields}
\end{figure}

\cref{fig:mnist-receptive-fields} compares the receptive fields of the infomorphic neurons to those obtained from logistic regression. Overall, we find a high degree of cosine similarity. Furthermore, visual inspection as well as the cosine similarity for each pair of corresponding rows and columns indicate that the receptive fields are particularly similar in the center of each image that contains the pixels that are relevant for classification~\cref{fig:mnist-receptive-fields}. Meanwhile, differences between the receptive fields are most pronounced closer to the image borders, with infomorphic networks seemingly utilizing these non-coding pixels as an additional bias term.

\begin{figure}[!ht]
    \centering
    \includegraphics[width=0.8\textwidth]{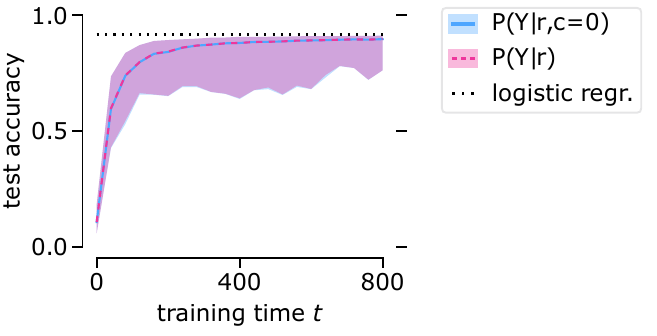}
    \caption{
    \textbf{The test accuracy of supervised infomorphic learning is similar using $P(Y\mid r,0)$ or $P(Y\mid r)$.} The average winner-take-all test accuracy using $P(Y\mid r,0)$ and $P(Y\mid r)$ across 100 training runs, with both accuracies approaching that of logistic regression (reaching on average $89.7\%$ vs. $91.9\%$  for log. regr.). Note that the 95-percentile is being displayed.
}
    \label{fig:mnist-optimal-accuracy}
\end{figure}

\cref{fig:mnist-optimal-accuracy} compares two approaches of computing the test accuracy: the \emph{absent-label approach} where the test accuracy is the winner take all accuracy $P(Y\mid r,0)$, i.e. $x_c=0$ during testing and the \emph{marginalized-label approach} where the test accuracy is the winner take all accuracy using $P(Y\mid r) \approx \sum_{c_\mathrm{train}}P(Y\mid r, c_\mathrm{train})P(c_\mathrm{train})$, i.e., $c$ is provided but marginalized using the prior probabilities $P(c_\mathrm{train})$ from the training set during testing. Thus, in different ways, both approaches hide the information of the actual contextual input (teacher signal) from the neuron while making the decision. Empirically, both approaches yield indistinguishable accuracy, highlighting that contextual inputs only have a modulating role, in line with our choice of activation function. Importantly, however, only the first approach is plausible in actual neural networks, as neurons cannot marginalize over their inputs before calculating their output.

\subsection{Unsupervised learning}
\begin{figure}[!ht]
    \centering
    \includegraphics[width=\textwidth]{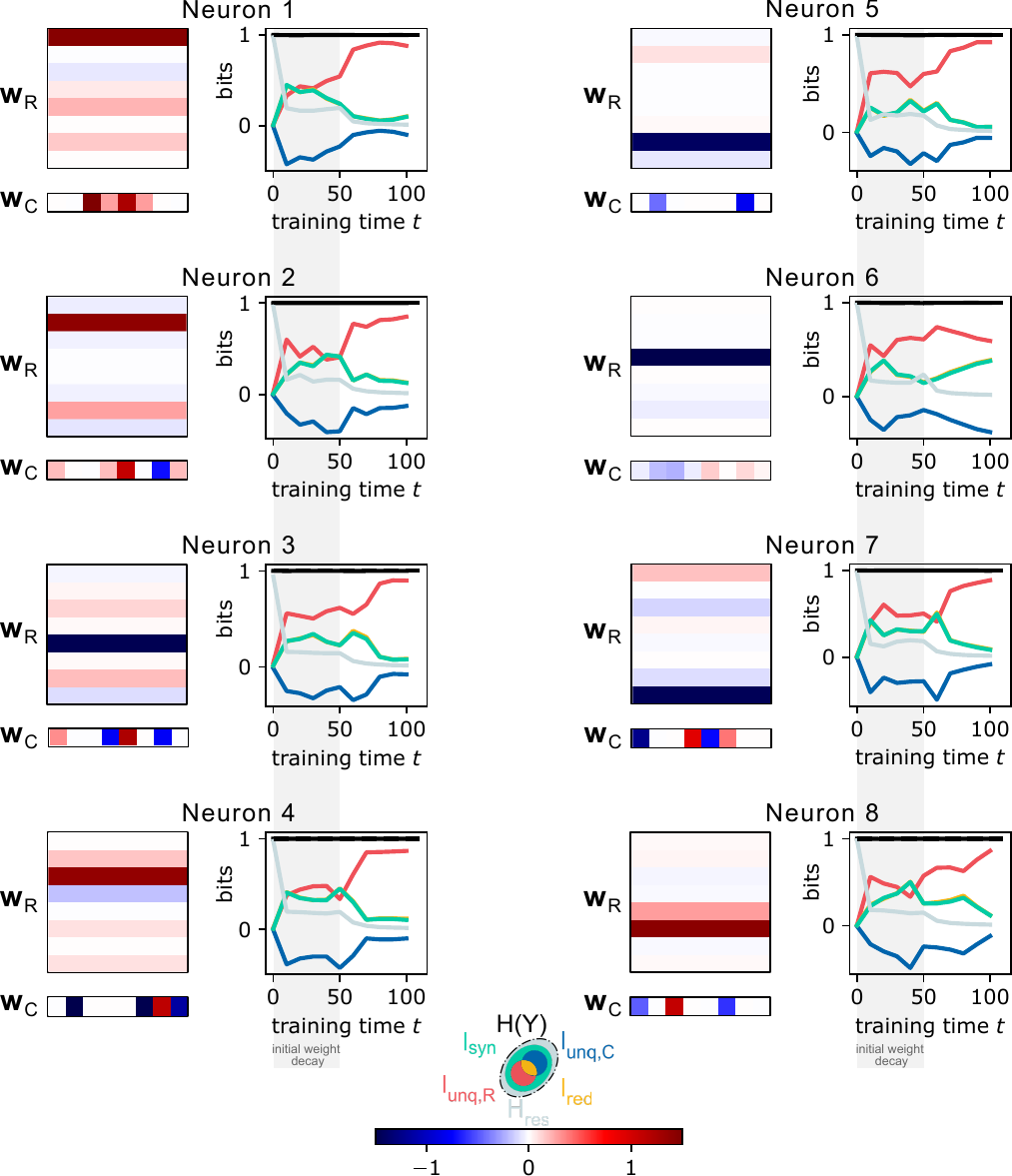}
    \caption{\textbf{Infomorphic neurons learn to encode distinct input features by unsupervised maximization of $I_{\unq,R}$ .} The receptive fields $\mbf w_R$ and contextual fields $\mbf w_C$ (first and third columns), and the evolution of all information contributions over learning (second and fourth columns) are shown for each neuron of a randomly chosen network \emph{successfully} performing unsupervised learning. The receptive input consists of 8 horizontal bars in an 8-by-8 grid, each bar appearing with probability $p = 0.5$. The contextual input is a vector of length 7, transmitting the output from all other neurons, without self-connections. Note that the values of all receptive and contextual weights are on the same scale indicated at the bottom of the figure.}
    \label{fig:bars-succ-pid-weights}
\end{figure}

\begin{figure}[!ht]
    \centering
    \includegraphics[width=0.8\textwidth]{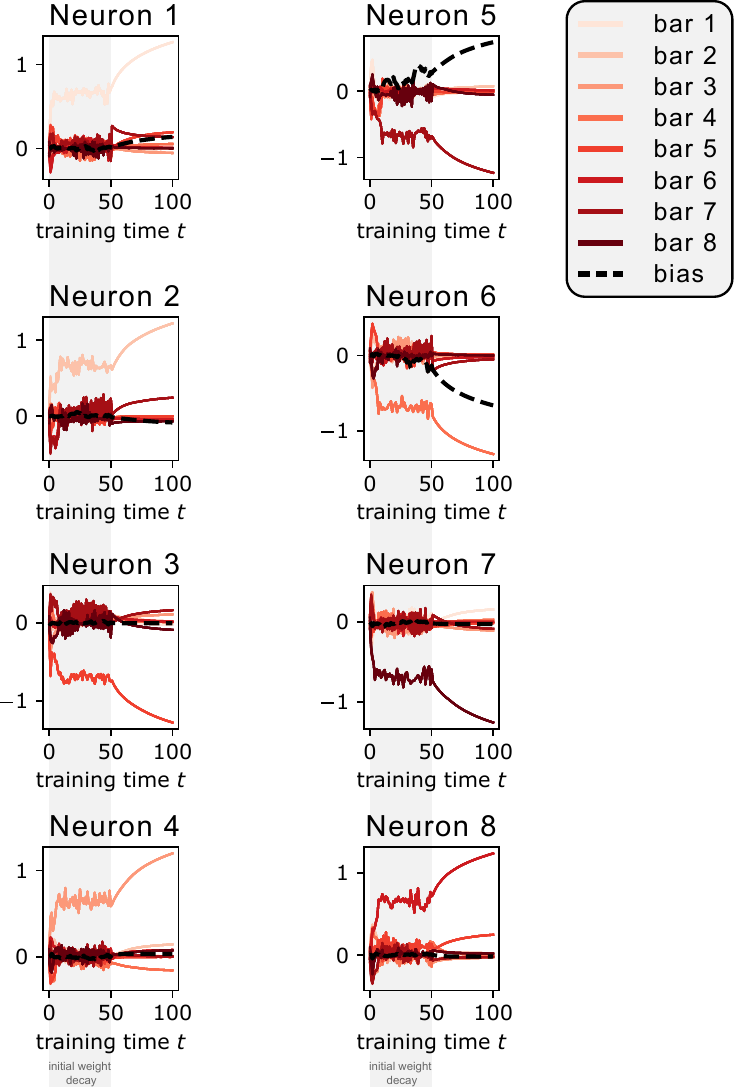}
    \caption{\textbf{Weight decay induces competition between neurons for individual bars.} The temporal evolution of the receptive weights is shown for each neuron of the same network as in~\cref{fig:bars-succ-pid-weights} (dashed black line is the bias). Each neuron receives 8 receptive inputs from each bar (one per pixel), yet as the pixel values are perfectly correlated, the 8 weights per neuron and bar follow the same gradient and almost perfectly overlap in the figure, where we plot them in the same color.}
    \label{fig:bars_succ_Wr_evol}
\end{figure}

\begin{figure}[!ht]
    \centering
    \includegraphics[width=0.8\textwidth]{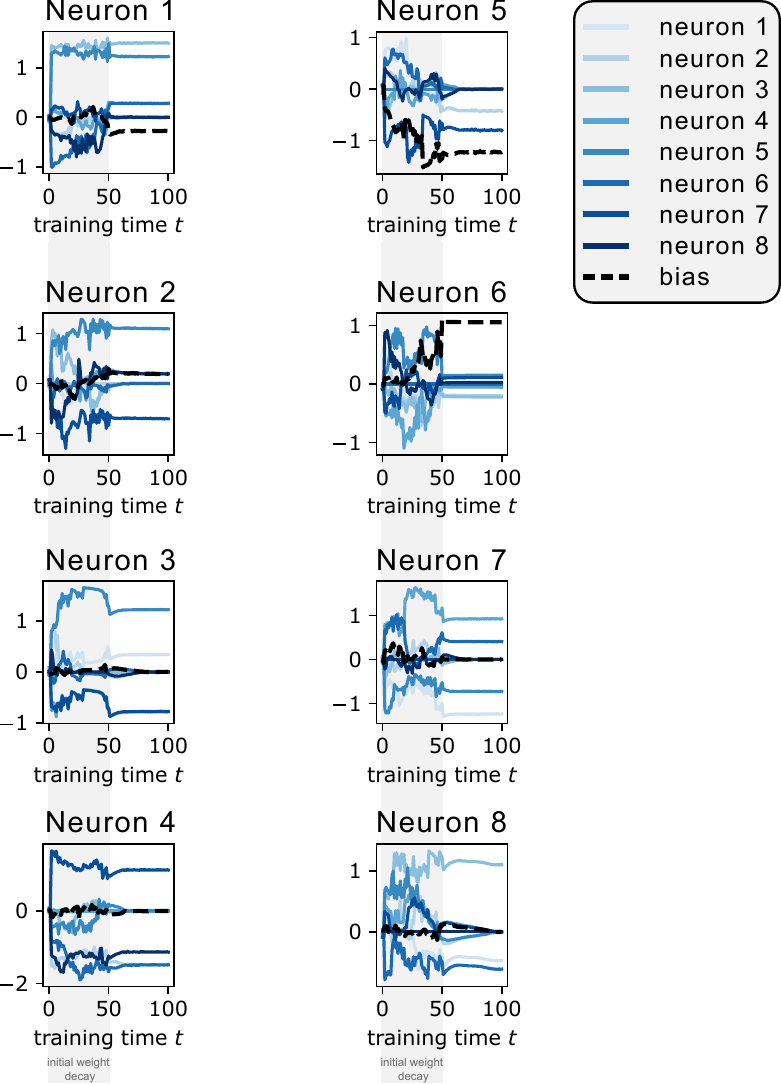}
    \caption{\textbf{Contextual weights almost exclusively change in the first phase of learning.} The temporal evolution of the contextual weights is shown for each neuron of the same network as in~\cref{fig:bars-succ-pid-weights} (dashed black line is the bias). There are no contextual self-connections in the network.}
    \label{fig:bars_succ_Wc_evol}
\end{figure}

\cref{fig:bars-succ-pid-weights} shows the evolution of the local information contributions of all the eight neurons over the course of training, and their receptive and contextual fields after training, from one randomly chosen network out of 298 networks that successfully encoded all eight bars. For the same network, \cref{fig:bars_succ_Wr_evol} shows the evolution of the receptive weights of all neurons, and \cref{fig:bars_succ_Wc_evol} shows the evolution of the contextual weights. It is evident from these three figures that all relevant learning happens in the first phase of training (specifically the first 20 mini-batches) which includes a weight decay on the receptive weights after every training step. Due to the weight decay, all neurons have overall low receptive weights, meaning their firing is highly stochastic. In this phase, neurons are competing for the encoding of individual bars, as seen from their receptive weights (\cref{fig:bars_succ_Wr_evol}), until all neurons have settled on an individual bar. This happens because coding for the same bars introduces redundant information $I_{\red}$, which neurons need to reduce in order to maximize their unique receptive information $I_{\unq, R}$, as by definition $I_{\red} + I_{\unq, R} = I(Y:R) \leq 1$.

In the second training phase, neurons mainly increase the receptive weight of their preferred bar, reducing their stochasticity and thereby increasing their unique information further. To our surprise, in this phase the contextual weights stay almost constant, an effect for which we currently have no intuitive explanation. Due to these constant and often non-zero contextual weights, each neuron is influenced to varying degrees by the other neurons. This influence leads each neuron to partly code for the bars of other neurons, despite our choice of an activation function that predominantly depends on the receptive input. This indirect encoding of bars other than the preferred one introduces a certain degree of redundant information $I_{\red}$, as the information about these non-preferred bars is also contained in the receptive input, i.e. the image. Interestingly, most neurons learn to compensate for this indirect influence by learning receptive weights for the respective bars that counteract the effect of these contextual inputs on their output. It is for this reason that the receptive fields of most neurons show additional weak traces of bars other than their preferred bar (\cref{fig:bars-succ-pid-weights}).

\begin{figure}[!ht]
    \centering
    \includegraphics[width=\textwidth]{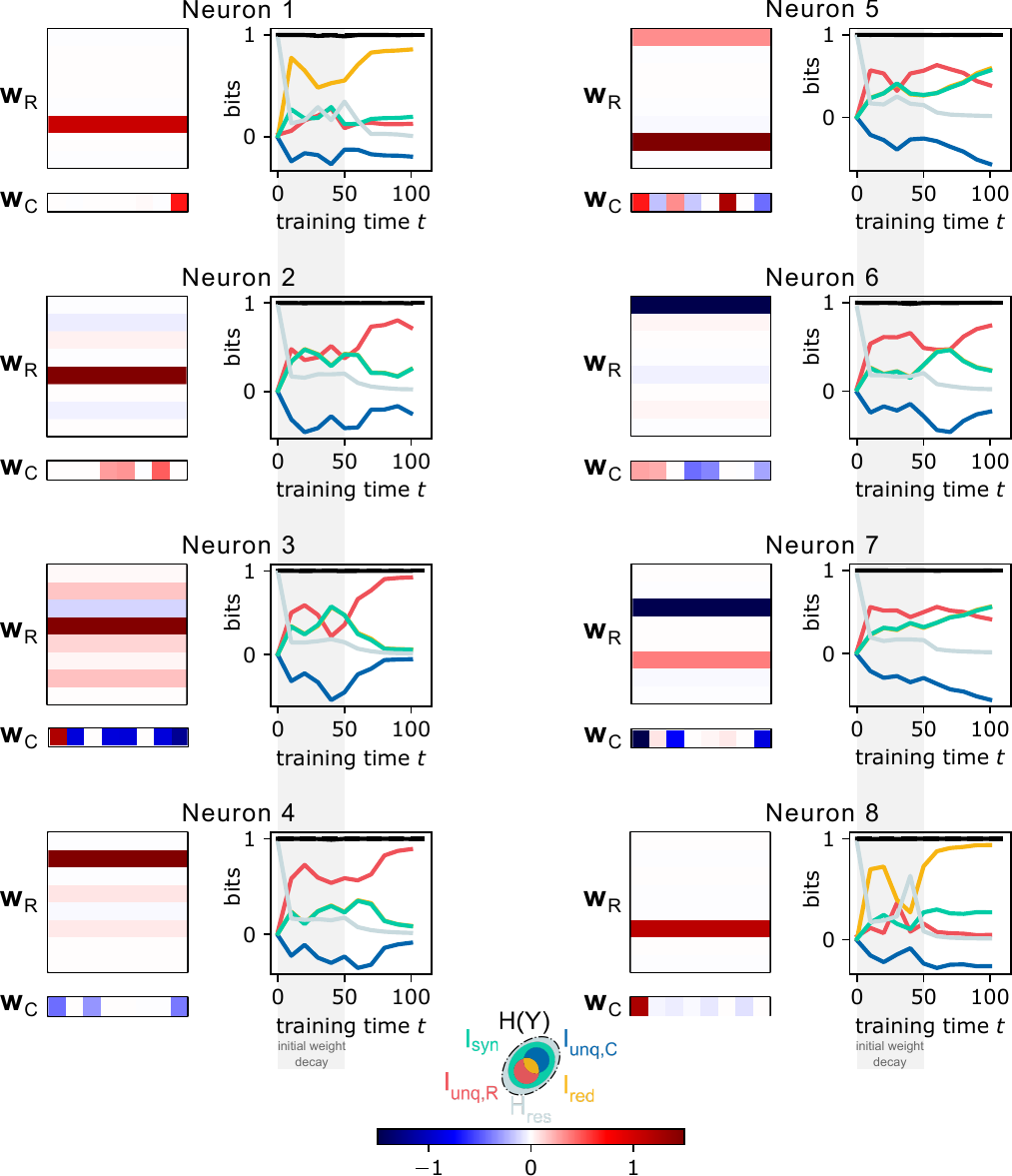}
    \caption{\textbf{Rare errors of two neurons encoding the same bar result in high redundant information $I_{\red}$.} The receptive fields $\mbf w_R$ and contextual fields $\mbf w_C$ (first and third columns), and the evolution of all information contributions over learning (second and fourth columns) are shown for each neuron of a randomly chosen network \emph{unsuccessfully} performing unsupervised learning. Note that this is one of the two networks that encoded only 7 out of the 8 available bars. Additionally note that all the receptive and contextual weight fields are on the same scale indicated at the bottom of the figure.
    }
    \label{fig:bars-unsucc-pid-weights}
\end{figure}

\cref{fig:bars-unsucc-pid-weights} shows the evolution of the local information contributions of all the eight neurons over the course of training, and their receptive and contextual fields after training, from one randomly chosen network out of 2 unsuccessful networks. Here, the first and the last neuron converged onto the same bar, while the last bar is not encoded by any neuron. As a result, both neurons fail to increase their unique receptive information $I_{\unq, R}$ and instead exhibit high redundant information $I_{\red}$ after training.

\begin{figure}
    \centering
    \includegraphics[width=\textwidth]{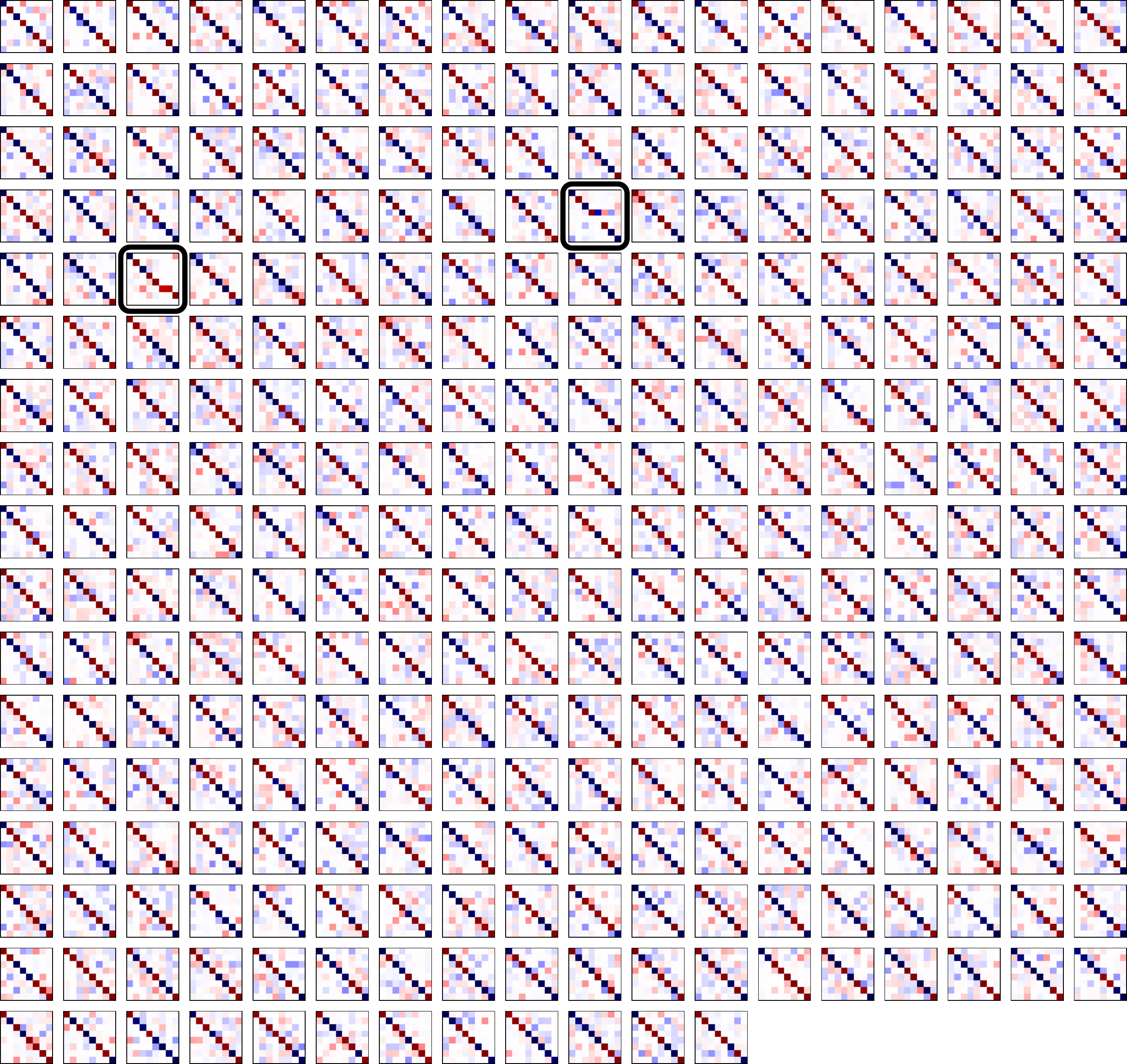}
    \caption{\textbf{Two out of 300 unsupervised infomorphic networks failed to encode all eight bars.} We show a compressed version of the receptive fields for all 300 networks. All eight receptive weights connecting the pixels of each bar to each neuron highly overlap, so we only report the weight of the left-most pixel from each bar to each neuron. This results in an 8-by-8 matrix summarizing the receptive fields of all neurons in a network (rows are bars, columns are neurons). We sort the neurons by their preferred bars, such that dark pixels along the diagonal indicate perfect encoding. The two networks with sub-optimal encoding are framed in black.}
    \label{fig:bars-encoding-all-nets}
\end{figure}

\cref{fig:bars-encoding-all-nets} shows a summary of the receptive weights from all 300 randomly initialized runs, out of which 298 led to each neuron encoding a separate bar, and 2 led to an encoding of only seven out of eight bars, with one redundantly encoded bar as shown in~\cref{fig:bars-unsucc-pid-weights}.

\subsection{Associative memory} 
\begin{figure}[!ht]
    \centering
    \includegraphics[width=\textwidth]{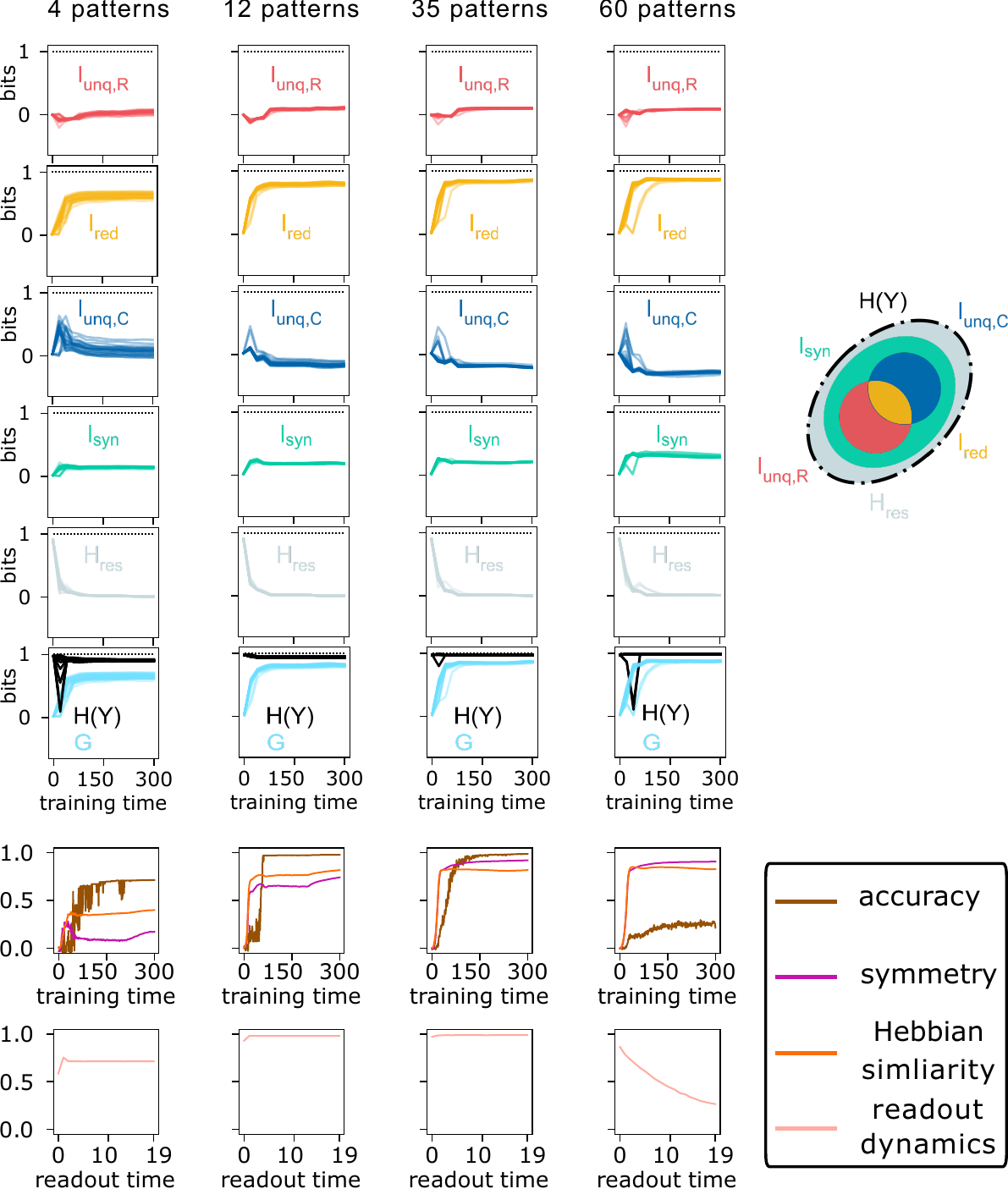}
    \caption{\textbf{Maximizing $I_{\red}$ leads to an asymmetric learning rule for auto-associative memories, which is different from outer-product Hebbian learning.} In the first six rows, we show the evolution of the PID information atoms $I_{\unq,R}, I_{\red}, I_{\unq,C}$ and $I_{\syn}$, the residual entropy $H(Y\mid R, C)$, the full output entropy $H(Y)$ and the goal function $G= 0.1I_{\unq,R} + 0.1I_{\unq,C} + I_{\red} + 0.1I_{\syn}$ over time. Each information quantity is averaged over all 100 neurons in a network, for each of 25 networks trained on either 4, 12, 35, and 60 patterns (columns). In the fifth row, we show for one randomly chosen networks of each size: the cosine similarity of the readout pattern with the correct memory pattern after 20 time steps (``accuracy''), the cosine similarity of the recurrent contextual weight matrix $W_C$ and its transpose (``symmetry''), and the cosine similarity of $W_C$ with the corresponding outer-product Hebbian weight matrix \citep{hopfield1982neural} (``Hebbian similarity'') over training time. In the sixth row, we plot for the same networks the cosine similarity of the readout pattern with the correct memory pattern over the 20 time steps of the readout. The cue pattern is presented only in the first time step, afterwards the receptive input is set to $\mbf x_R = 0$.}
    \label{fig:memory-appendix}
\end{figure}

\cref{fig:memory-appendix} shows the local information contributions averaged over all neurons, for 25 networks trained on 4, 12, 35, 60 patterns each.
As mentioned in the main text, infomorphic networks cannot reliably encode very few (e.g.\ 4) patterns, as in this case $\frac{100}{2^3}=12.5$ neurons are expected to receive the exact same receptive input for every pattern, resulting in $I(Y: R) =0$ bits of information in their receptive inputs. Concurrently, the average redundant information $I_{\red}$ is comparably low and the unique contextual information $I_{\unq, C}$ comparably high in these networks, indicating that the recurrent weights of certain neurons are not aligned with their external input, but instead retain unique information induced by their random initialization.

All other network sizes show high redundant information $I_{\red}$, despite the network with 60 patterns exhibiting low readout accuracy, taken as a sign of being above capacity. The high redundant information even above the capacity limit is compensated by unique contextual misinformation, i.e. $I_{\unq, C}<0$. Negative information terms are possible in the $I^{\sx}_\cap$ measure, however, the consistency equation $I(Y: C) = I_{\red} + I_{\unq, C}$ must still be satisfied. This results in low mutual information $I(Y: C)$ despite high redundant information $I_{\red}$, indicating that the recurrent weights are in general not sufficient to predict the activity of the neuron in these networks, and thus firing patterns cannot be reliably sustained without external input.

Given that infomorphic networks exhibit a higher capacity than classical Hopfield networks trained with one-shot outer-product Hebbian learning (as shown in the main text), we calculate two further measures to compare these learning rules. Firstly, we calculate the cosine similarity between the weight matrix obtained from infomorphic networks and the weight matrix obtained from outer-product Hebbian learning on the same memory patterns. Given that outer-product Hebbian learning leads to a symmetric weight matrix, we secondly calculate the symmetry of the obtained infomorphic weight matrix, as the cosine distance between the weight matrix and its transpose. We find consistent deviations both from perfect similarity and from perfect symmetry, which seem to get smaller for larger training sets (\cref{fig:memory-appendix}). Further investigation of this phenomenon and comparisons to previously published online learning rules for associative learning are left for future work.

Finally, readout accuracy increases over training time, and networks with 12 and 35 patterns show almost instantaneous convergence to the correct memory pattern over readout time (\cref{fig:memory-appendix}). In contrast, networks with 4 patterns show stable readout accuracy below 1, hinting at a stable sub-network that reaches a fixed point, in line with the idea that some neurons cannot learn due to lack of information in their receptive inputs. On the contrary, the readout accuracy for networks with 60 patterns decreases gradually over readout time, indicating an inability to sustain the cued pattern. Note that we do not binarize our readouts but keep the stochastic sampling, which leads to an unfair comparison, even impeding performance of the infomorphic networks in comparison to classical Hopfield networks.

\section{Supplementary Experiments\label{apx:suppl-exp}}

\subsection{Primitive error neurons \label{apx:error-neurons}}
In this section we construct simple ``error neurons'', as might later be used in infomorphic approaches to predictive coding~\citep{rao1999predictive, millidge2022predictive}.

\begin{figure}[!ht]
    \centering
    \includegraphics[width = \textwidth]{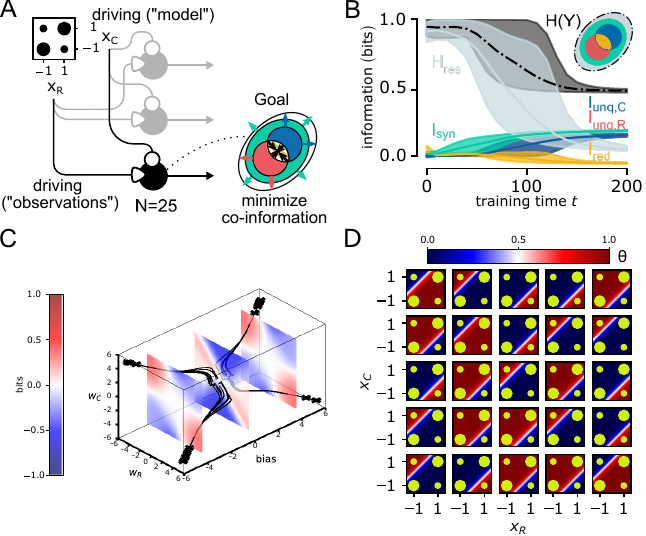}
    \caption{\textbf{Binary infomorphic error neurons via maximization of synergy and uniqueness and minimization of redundancy.} By maximizing $G = I_{\syn} + I_{\unq, R} + I_{\unq, C} - I_{\red}$ between two binary, correlated inputs, infomorphic neurons learn to code for rare events, resulting in signed error neurons. (A) Network architecture for a set of neurons that do not communicate laterally, each receiving the same scalar binary inputs $\mbf X_R$ (the ``observations'') and $\mbf X_C$ (the ``model''). The inputs are both Bernoulli-distributed with $p = p(X_R=1) = p(X_C=1) = .5$, and correlated with $p(X_R=X_C=1)=p(X_R=X_C=-1)=.4$ and $p(X_R=-X_C=1)=p(-X_R=X_C=1)=.1$. Choosing the symmetric activation function $A(r, c) := r + c$ gives equal importance to model and observations, and results in a single bias term of the form $\mathrm{bias} = w_{0,R} + w_{0,C}$. (B) Information quantities and goal function averaged over all 25 independently trained neurons. Shaded areas correspond to the 95-percentile. (C) Graphical depiction of the full learning dynamics. Black lines show trajectories in parameter space over the course of learning, with the endpoint of learning indicated by a ``+'' marker. Depending on their random initialization, neurons pick one of four corners of the three-dimensional parameter cuboid. Colored surfaces show the goal function $G$ for different values of the weights at fixed bias. (D) Boundaries in input space for all 25 neurons after training. Background color indicates firing probability $\theta = p(Y=1\mid X_R,X_C)$, green disks denote the input distribution, with the area of a disk corresponding to the probability of the respective event (compare with panel A). The neurons randomly choose one of the two rare events and learn to separate it from all other events. Taking multiple error neurons into account, a downstream neuron could use this as a signed error indicator detecting disagreement between the model and the observations.}
    \label{fig:error}
\end{figure}

\paragraph{Topology and Inputs}
We train 25 infomorphic neurons with correlated scalar binary inputs as in the previous section, i.e., the receptive and contextual inputs to the neuron are jointly distributed as a two-dimensional Bernoulli distribution $p(X_R,X_C)$ with $p(X_R=-1,X_C=-1) = p(X_R=1,X_C=1) = .4$ and $p(X_R=-1,X_C=-1) = p(X_R=1,X_C=1) = .1$, resulting in a Pearson correlation of $.6$. As in the supervised learning experiments from the main text, the neurons are not connected to each other, as that would require a third input that is currently lacking (\cref{fig:error} A).

\paragraph{Goal Functions}
We set the goal function of all neurons to $G = I_{\syn} + I_{\unq, R} + I_{\unq, C} - I_{\red}$. Intuitively, a perfect model would have maximal mutual information with the observations, such that an error neuron could then only extract redundant information between prediction and observation. As a perfect model should not be changed, in this case an optimal error neuron should have zero entropy. To achieve that, error neurons need to explicitly minimize redundant information $I_{\red}$. All other information atoms correspond to a mismatch between model and observation, so should be maximized. Specifically, maximizing conditional information of the receptive input $I(Y:R\mid C) = I_{\syn} + I_{\unq, R}$ makes the neuron encode for all information that is observed but has not been predicted. Simultaneously maximizing contextual unique information $I_{\unq, C}$ makes the neuron encode all information that has been predicted but not observed, thus providing a pruning signal to the model to not make unnecessary predictions.

\paragraph{Activation Functions}
We choose a simple symmetric activation function $A(r,c) = r+c$ because errors neurons should be driven by either input. Also this allows us to reduce each neuron to three trainable parameters $w_R$, $w_C$ and $\mathrm{bias} = w_{R,0} + w_{C,0}$, such that the full learning trajectories can be graphically depicted.

\paragraph{Protocol}
We train each neuron on $200$ batches of $1000$ randomly sampled stimuli. After each batch, we update the parameters (\cref{tab:err} for all chosen training parameters).

\paragraph{Performance and Outcome}
Neurons learn to code for one of the infrequent events in the distribution, i.e. either $X_R = -1, X_C = 1$ or $X_R = 1, X_C = -1$, at random (\cref{fig:error} D). This is intuitively correct, as these infrequent events signal sub-optimal prediction: An optimal binary prediction would perfectly correlate to the binary observation, such that $p(X_R = -1, X_C = 1) = p(X_R = 1, X_C = -1) = 0$. Note that an infomorphic neuron with a linear decision boundary cannot encode both infrequent events. However, a downstream prediction neuron might listen to multiple binary error neurons to get a full signed prediction error. We speculate that a ternary neuron with an output alphabet of three elements could be trained in a similar manner to code for both infrequent events. Alternatively, infomorphic neurons with three input classes could coordinate laterally, as in the unsupervised learning experiments. We furthermore speculate that such three-input-classes infomorphic neurons could also be used to signal prediction errors when the observations and model predictions are more high-dimensional and errors can exceed one bit of information.

\paragraph{Information Dynamics}
As expected, synergistic and unique information atoms increase over learning, and redundant information decreases. Furthermore, the neurons become less stochastic, as seen by a decrease of residual entropy. Importantly, the entropy $H(Y)$ of the error neurons decreases over learning, indicating that the amount of encodable ``error information'' is smaller than one bit, and that the neurons are indeed not encoding any redundant information or noise (\cref{fig:error} B and C).

\begin{table}[!ht]
    \caption{\label{tab:err}\textit{Parameters of the infomorphic neurons in the Primitive error neurons experiment}.} 
    \centering
    \begin{tabular}{|c|c||c|}
        \toprule
        \multicolumn{3}{|c|}{Training}\\
        \cmidrule{1-3}
        \multicolumn{1}{|c|}{Parameter}    &\multicolumn{1}{c||}{Value}    &\multicolumn{1}{c|}{Comment}\\
            \cmidrule(r){1-2}\cmidrule(lr){3-3}
            Phases & 200 & a single phase of training with 200 epochs\\
            $N_\mathrm{tr}$ & 1000& sampled from the set of $x,z\in\{-1,1\}$\\
            $N_\mathrm{te}$ & 10000& sampled from the set of $x,z\in\{-1,1\}$\\
            $m_{\mathrm{rep}}$ & 1 & no repetition is needed\\
            \cmidrule{1-3}\morecmidrules\cmidrule{1-3}
        \multicolumn{3}{|c|}{Learning}\\ 
        \cmidrule{1-3}
        \multicolumn{1}{|c|}{Parameter}    &\multicolumn{1}{c||}{Value}    &\multicolumn{1}{c|}{Comment}\\
            \cmidrule(r){1-2}\cmidrule(lr){3-3}
            $b_\mathrm{init}$ & 0.5 & all weights and biases initialized in $[-0.5,-0.125]\cup[0.125,0.5]$,\\
             & & no qualitative difference for initialization in $[-0.5,0.5]$\\
            $\mbf \Gamma$ &(1, 1, -1, 1, 0) & goal parameters ($\Gamma_{\unq, R}$, $\Gamma_{\unq, C}$, $\Gamma_{\red}$, $\Gamma_{\syn}$, $\Gamma_{\res}$)\\
            $\eta$ & 1 & learning rate \\
            $\lambda$ & 0.0 & no weight pullback\\
            \cmidrule{1-3}\morecmidrules\cmidrule{1-3}
        \multicolumn{3}{|c|}{Input Integration}\\ 
        \cmidrule{1-3}
        \multicolumn{1}{|c|}{Parameter}    &\multicolumn{1}{c||}{Value}    &\multicolumn{1}{c|}{Comment}\\
            \cmidrule(r){1-2}\cmidrule(lr){3-3}
            $n_{\mathrm{receptive}}$ & 1 & $x\in\{1,-1\}$\\
            $n_{\mathrm{contextual}}$ & 1& $z\in\{1,-1\}$\\
            $J_R$ & $[-8, 8]$ & different range does not affect the training\\
            $J_C$ & $[-8, 8]$ & different range does not affect the training\\
            $n_R$-bins & $500$ & uniform bin-size is $0.032$\\
            $n_C$-bins & $500$ & uniform bin-size is $0.032$\\
            \bottomrule
        \end{tabular}
    \end{table}

\end{document}